\documentclass{article}

\usepackage{bm}
\usepackage{graphicx}
\usepackage{amscd,amsmath,amssymb,amsthm}
\usepackage{lscape}
\usepackage{color}

\newcommand{\be}{\begin{equation}}
\newcommand{\bel}{\begin{equation}\label}
\newcommand{\ee}{\end{equation}}
\newcommand{\bc}{\begin{center}}
\newcommand{\ec}{\end{center}}
\begin{document}

\title{Self-similar source-type solutions to the three-dimensional Navier-Stokes equations}

\author{
  K. Ohkitani\\
  Research Institute for Mathematical Sciences,\\
  Kyoto University,  Kyoto 606-8502 Japan.\\
\and
R. Vanon\\
School of Mathematics, Statistics and Physics, \\
    Newcastle University, Newcastle upon Tyne, NE1 7RU,  U.K.
}




\maketitle
\begin{abstract}
We formalise a systematic method of constructing
forward self-similar solutions to the Navier-Stokes equations
in order to characterise the late stage of decaying process of turbulent flows.
  (i) In view of critical scale-invariance of type 2 we exploit the vorticity curl as the dependent variable
  to derive and analyse the dynamically-scaled Navier-Stokes equations. This formalism offers
  the viewpoint from which the problem takes the simplest possible form.
  (ii) Rewriting the scaled Navier-Stokes equations by Duhamel principle as integral equations, we regard the nonlinear term as
  a perturbation using the Fokker-Planck evolution semigroup.
  Systematic successive approximations are introduced  and the leading-order solution
  is worked out explicitly as the Gaussian function with a solenoidal projection.
  (iii)  By iterations the second-order approximation is  estimated explicitly up to  solenoidal
  projection and is evaluated numerically.
  (iv)  A new characterisation of nonlinear term is introduced on this  basis to estimate its  strength $N$
  quantitatively. We find that  $N=O(10^{-2})$ for the 3D Navier-Stokes equations. This should be contrasted with
$N=O(10^{-1})$ for the Burgers equations and $N \equiv 0$ for the 2D Navier-Stokes equations.
(v) As an illustration we explicitly determine source-type solutions to the  multi-dimensional
  the Burgers equations. Implications and applications of the current results are given.


\end{abstract}
\begin{flushleft}
Navier-Stokes equations,self-similarity, scale-invariance
\end{flushleft}


\section{Introduction}
\textcolor{black}{Self-similarity is a tool of fundamental importance in analysing partial differential equations,
  including the construction of solutions and the determination of their stability. It is also useful
  for numerical and asymptotic methods of studying  partial differential equations.
  For general aspects of self-similarity and its applications we refer the readers  to e.g.
  \cite{Barenblatt2003,Dresner1983}.} 

\textcolor{black}{In this paper} we will study the so-called source-type self-similar solution to the Navier-Stokes
equations. Our motivation for this is as follows.
First of all, it will give us a
particular self-similar solution to the Navier-Stokes equations, which
characterises the decaying process in the late stage of evolution. Second,
it is likely to give useful information as to how we may handle more general
solutions.

\textcolor{black}{It may be in order to have a look at previous works which are related to this paper.} 
It has been shown under mild conditions that no nontrivial smooth backward self-similar
solution exists to the Navier-Stokes equations. On the other hand it is known that
nontrivial forward self-similar solutions {\it do} exist, but their explicit
functional forms are not known, except for some asymptotic results. 
It is of interest to see \textcolor{black}{how} they actually behave because
such solutions contain  important information regarding more general
solutions.
This is particularly the case when the governing equations are exactly
linearisable, e.g. the Burgers equations. While it is not expected
that the Navier-Stokes equations are exactly soluble in general, we might still obtain
insights into the nature of their solutions.

In \cite{CP1996}  the existence of forward self-similar solutions
for small data was proven by using a fixed-point theorem in a Besov space (see below).
There, initial data for the self-similar solution (3D)
\textcolor{black}{are} assumed to be homogeneous of degree $-1$ in velocity
$$\bm{u}_0(\lambda \bm{x})=\lambda^{-1} \bm{u}_0(\bm{x})$$
and the existence of a self-similar solution of the form
$$\bm{u}(\bm{x},t)=\frac{1}{\sqrt{t}}\bm{U}\left(\frac{\bm{x}}{\sqrt{t}} \right)$$
has been established under the assumption that initial data is small in some Besov space.
Moreover, it has been proved that the self-similar profile $U$ satisfies (in their notations)
$$U=S(1) u_0 +W,$$
where $S(1)$ denotes a heat operator at time 1 and $\|W\|_{L^3}$ is small.
In \cite{JS2014} using a locally H{\"o}lder class in $\mathbb{R}^3\backslash\{0\}$
the smallness assumption has been removed and
it is furthermore shown that
$$|U(x)-e^{\triangle}u_0(x)| \leq \frac{C(M)}{(1+|x|)^{1+\alpha}},$$
where $0 < \alpha <1$ and $C(M)$ denotes a constant with  some norm $M$ of $u_0$.   
Those studies indicate that the self-similar solution is close to the heat flow in the late
stage. However, studies on the determination of a specific functional form of
self-similar solutions are  few and far between, except for an attempt in\cite{Brandolese2009}.
We also note that the existence of generalised self-similar solutions (in the sense that the scaling holds
only at a set of discrete values of $\lambda$) was studied subsequently, e.g. \cite{CW2017, BT2019}.

\textcolor{black}{
Our basic strategy in practice is as follows.  After recasting the dynamically-scaled Navier-Stokes equations
as integral equations via the Duhamel principle, we regard the nonlinear term as a perturbation using the Fokker-Planck
evolution semigroup.
  Systematic successive approximations are then introduced  and the first-order solution
  is worked out explicitly as the Gaussian function with a solenoidal projection. The second-order
approximation is also evaluated numerically to assess the strength of the nonlinear term.}

We will  construct an approximate solution to the 3D Navier-Stokes equations valid in the long-time limit, using
the vorticity curl \textcolor{black}{$\nabla \times \bm{\omega}$}.
In Sections 4 we will see why this is the most convenient variable from the reaction of
the Navier-Stokes equations under dynamic scaling.

\textcolor{black}{Here we appreciate the suitability of such a choice of the unknown
  by comparing the source-type solutions to the Navier-Stokes equations
  and their self-similar solutions at criticality.
By definition the source-type solution for nonlinear parabolic PDEs is a solution in a scaled space,
which \textit{starts from the Dirac mass} in some dependent variable and \textit{ends up like a near-identity of
the Gaussian function} in the long-time limit. It serves as an analogue of the fundamental solution to nonlinear PDEs.}

The unknown whose $L^1$-norm is marginally divergent is suitable for describing the late-stage
evolution. This is because this self-similar solution satisfies the same scaling as the Dirac mass and both of them
belong to a Besov space near $L^1$. 
In one dimension, in the limit of $t \to 0$, we have roughly
$$u \sim \dfrac{1}{x},\; \mbox{marginally}\,\notin L^1(\mathbb{R}^1),$$
which \textcolor{black}{suggests} that the velocity is convenient in this case.

In two dimensions it is the vorticity which is the most convenient, 
as can be seen from
$$\omega \sim \dfrac{1}{r^2},\;\mbox{marginally}\,\notin L^1(\mathbb{R}^2),$$
where $|\bm{x}|=r$.
Recall that those scaling properties of velocity in 1D or vorticity in 2D,
are the same as that of the Dirac mass;
$\lambda^d \delta(\lambda \bm{x})=\delta(\bm{x})$
in $d$-dimensions.

Now consider Besov spaces whose norms are given by
$$\|\bm{u}\|_{B^s_{pq}} \equiv \left\{\sum_{j=1}^\infty
\left( 2^{sj} \|\Delta_j (\bm{u}) \|_{L^p}\right)^q \right\}^{1/q},$$
where $1 \leq p,q \leq \infty, s \in \mathbb{R}$ and $\Delta_j (\bm{u})$ represents band-filtered velocity
at frequency $2^j$.
It is known that in $\mathbb{R}^d$ the Dirac delta mass is embedded  as
\bel{delta}
\delta(\bm{x}) \in B_{p,\infty}^{-d+d/p},\;\;\mbox{for}\;\; p \geq 1,
\ee
see e.g. \cite{HL2017}. In particular we have
$\delta(\bm{x}) \in B_{1,\infty}^{0}$ for any $d.$

While the velocity $u \sim \dfrac{1}{r} \notin L^3(\mathbb{R}^3),$ we have 
$$u \sim \frac{1}{r} \in B_{3,\infty}^{0}(\mathbb{R}^3),$$ and correspondingly for $\bm{\chi}=\nabla \times \bm{\omega},$
the vorticity curl,
$$\chi \sim \frac{1}{r^3} \in B_{1,\infty}^{0}(\mathbb{R}^3).$$ 
Hence in three dimensions this $\chi$ and the Dirac mass belong to the same
function class $B_{1,\infty}^{0}(\mathbb{R}^3)$, with $p=1$ in (\ref{delta}).

\textcolor{black}{The rest of this paper is organised as follows. In Section 2 after reviewing
  critical scale-invariance of type 2 (to be defined below) using the Burgers equation we introduce successive approximations
  of determining the self-similar profile. On this basis we introduce and quantify the strength of nonlinearity.
  Higher-dimensional Burgers equations are also discussed. In Section 3 we have a brief look at the 2D
  Navier-Stokes equations. In the main Section 4  we study the self-similar solutions of the 3D Navier-Stokes
  equations utilising the vorticity curl as the unknown to achieve critical scale-invariance of type 2.
  We carry out successive approximations and determine the strength of nonlinearity as introduced above. Section 5
  will be devoted to the Summary and outlook. Some further details and derivations are given in Appendices.}

\section{Burgers equations}
\textcolor{black}{
  {\it We review the source-type solution of the Burgers equation with an emphasis on the critical scale-invariance.
    Our approach is novel in the introduction of its successive approximations, in preparation for handling the 3D Navier-Stokes
    equations, and in employing a new method of estimating the strength of nonlinearity on this basis.
    Also described are     the source-type solutions of the Burgers equations in $n$-dimensions. }}
\subsection{\textcolor{black}{Critical scale-invariance}}
We consider the Burgers equation \cite{Burgers1948} 
\bel{1DBurgers}
\frac{\partial u}{\partial t} + u \frac{\partial u}{\partial x}
=\nu \frac{\partial^2 u}{\partial x^2},
\ee
which satisfies static scale-invariance under
$$x \to \lambda x, t \to \lambda^2 t, u \to \lambda^{-1} u.$$
This means that if $u(x,t)$ is a solution, so is
\textcolor{black}{$u_\lambda(x,t) \equiv \lambda u(\lambda x, \lambda^2 t),$}
for any $\lambda (>0)$.
It is readily checked that 
$$\|u_\lambda\|_{L^p}=\lambda^{\frac{p-1}{p}}\|u\|_{L^p},$$
which shows that the $L^1$-norm is scale-invariant.

\textcolor{black}{
Let us clarify the two kinds of critical scale-invariance.
Type 1 scale-invariance  is achieved when we use a dependent  variable whose physical dimension is the same as $\nu$.
Type 1 is deterministic in nature where the additional term arising in the governing equations
under dynamic scaling is minimised in number. Type 2 instead is statistical in nature where the
additional terms under dynamic scaling are maximised in number so that
a divergence form is completed and the dynamically-scaled equations have the Fokker-Planck operator as the linearisation.}
In the former the dependent variable has the same physical dimension
as kinematic viscosity, whereas in the latter the argument of the Hopf
characteristic functional (the independent variable) has the same physical 
dimension as the reciprocal of kinematic viscosity \cite{Ohkitani2020}.
This approach provides a viewpoint from which the problem
appears in the simplest possible form.

Critical scale-invariance \textcolor{black}{of type 1} is achieved with the velocity potential $\phi,$ which is defined by
$u=\partial_x \phi$. If $\phi(x, t)$ is a solution, so is $\phi(\lambda x, \lambda^2 t).$
Under dynamic scaling for the velocity potential $\phi(x,t)=\Phi(\xi,\tau),$
\textcolor{black}{$\xi=\frac{x}{\sqrt{2 at}},\; \tau=\frac{1}{2 a}\log t$} we have
\bel{ScaledBurgers0}
\frac{\partial \Phi}{\partial \tau}
 + \frac{1}{2}\left(\frac{\partial \Phi}{\partial \xi}\right)^2
 =a  \xi\frac{\partial \Phi}{\partial \xi}+\nu \frac{\partial^2 \Phi}{\partial \xi^2},
 \ee
whose linearisation has the Ornstein-Uhlenbeck operator.
This is called type 1 (deterministic) scale-invariance where the number of additional terms is
minimised, that is, only the drift term remains.
Under dynamic scaling for velocity
\textcolor{black}{$u(x,t)=\frac{1}{\sqrt{2at}}U(\xi,\tau)$,}
we find
\bel{ScaledBurgers}
\frac{\partial U}{\partial \tau}
 +U \frac{\partial U}{\partial \xi}
=a  \frac{\partial}{\partial \xi}\left( \xi U \right)
+\nu \frac{\partial^2 U}{\partial \xi^2},
\ee
whose linearisation  is the Fokker-Planck equation. \textcolor{black}{Here the zooming-in parameter $a (>0)$
has the same physical dimension as $\nu$ and is on the same order of it.} With this  type 2 (statistical)
scale-invariance where the number of additional terms is maximised
meaning that  a divergence form is completed with the addition of $aU$ term.
As it is a second-order equation it has two independent
solutions, of which we will focus on the Gaussian one.
See  \textbf{Appendix C} for the other non-Gaussian kinds of solutions.

Equation (\ref{ScaledBurgers}) is exactly soluble and  its steady solution is called the source-type solution
\cite{BKW1999, BKW2001}:
\bel{Burgers_source1}
U(\xi)=\frac{U(0) \displaystyle{\exp \left( -\frac{a \xi^2}{2 \nu} \right)}}
{1 -\displaystyle{\frac{U(0)}{2\nu}\int_0^{\xi}} 
  \exp \left( -\frac{a \eta^2}{2 \nu}\right) d\eta}.
\ee
The name has come from the time zero asymptotics 
$$\lim_{t \to 0}\frac{1}{\sqrt{2at}} U(\xi)=M\delta(x),$$
where $M\equiv\int_{\mathbb{R}^1} u_0(x) dx$ and
$U(0)=\sqrt{\frac{8a\nu}{\pi}}\tanh\frac{M}{4\nu}
\approx \sqrt{\frac{a}{2\pi \nu}}M\;\; (\mbox{for}\;M/\nu \ll 1).$
Observe that (\ref{Burgers_source1}) is a \textit{near-identity} transformation of the Gaussian function.
See \cite{EZ1991, BKW1999, BKW2001}.

It is also known that for
$u_0 \in L^1$  we have 
$$t^{\frac{1}{2}\left(1-\frac{1}{p}\right)}
\left\| u(x,t)- \frac{1}{\sqrt{2at}} U(\xi)\right\|_{L^p}
\to 0\;\;\mbox{as}\;\; t \to \infty,$$
where $1 \leq p \leq \infty.$

The simplest method for solving (\ref{ScaledBurgers}) without linearisation
is as follows. Rewrite the equation 
\begin{eqnarray}
  \frac{U^2}{2}&=&a\xi U +\nu \frac{dU}{d\xi}  \nonumber\\
 &=&\nu \exp \left(-\frac{a \xi^2}{2\nu}\right)
  \frac{d}{d\xi}     \left(U \exp \left( \frac{a \xi^2}{2\nu} \right)\right).
 \nonumber
\end{eqnarray}
By changing variables to
$\tilde{U}=U \exp \left( \frac{a \xi^2}{2\nu}\right),\eta = \frac{1}{2\nu}
\int_0^\xi \exp \left(-\frac{a \zeta^2}{2\nu} \right) d\zeta,$ we find
$$\frac{d \tilde{U}}{d \eta}=\tilde{U}^2,$$
which is readily integrable. Alternatively we may solve the equation (2.1) by regarding it as a
Bernoulli equation.

It may be in order to comment on the significance of source-type solution.
When we recast (\ref{Burgers_source1}) as
\bel{Burgers_source2}
U(\xi)=-2\nu \frac{\partial}{\partial \xi}\log\left(
{1 -\displaystyle{\frac{U(0)}{2\nu}\int_0^{\xi}}   \exp \left( -\frac{a \eta^2}{2 \nu}\right) d\eta}
\right),
\ee
which is reminiscent of the celebrated Cole-Hopf transform. In other words, the source-type solution
encodes the vital information of the nonlinear term in the case of the Burgers equation. Note that the error-function itself
$\int_0^{\xi}   \exp \left( -\frac{a \eta^2}{2 \nu}\right) d\eta$  \textcolor{black}{in (\ref{Burgers_source1}, \ref{Burgers_source2})}
is a self-similar solution to
the heat equation. This suggests that studying source-type solution of the Navier-Stokes equations may give
a hint \textcolor{black}{on how to characterise their long-time evolution by a heat flow.}

\subsection{Successive approximations}
The operator $L = \triangle^* \equiv \triangle  +\frac{a}{\nu}\partial_\xi(\xi \cdot )
$ is not self-adjoint. It is possible to find a function $G$\footnote{With a slight abuse of notation
this  $G$ should be distinguished from the Gaussian function  used in Section 3.}
such that $L^{\dagger} G(\xi) =-\delta(\xi)$ holds, where
$L^{\dagger}\equiv \partial_\xi^2 -\frac{a}{\nu}\xi \partial_\xi$ is the adjoint of $L$.
In fact $G(\xi) \propto D\left( \sqrt{\frac{a}{2\nu}} \xi\right),$ where $D(\cdot)$ denotes
the Dawson's integral, \textcolor{black}{defined by $D(x) \equiv  e^{-x^2}\int_0^x e^{y^2} dy$  \cite{OLB2010}.}
However, because $G$ decays slowly at large distances
$G(\xi) \propto 1/\xi$ as $|\xi| \to \infty$, it cannot be used as a Green's function,
at least, in the usual manner.

The inversion formula for $\triangle^*$ can be obtained by an alternative method.
Recall that \textcolor{black}{based on} $\frac{1}{a}=\int_0^\infty e^{-at} dt \; (a>0),$
the fundamental solution to the Poisson equation in 1D is given by
$$ (\nu \triangle)^{-1} \equiv  -\int_{0}^{\infty} ds e^{\nu s \triangle} =\frac{|\xi|}{2\nu}*,$$
where * denotes convolution.
Likewise for the fundamental solution to the Fokker-Planck equation in 1D we write
$$(\nu \triangle^*)^{-1} \equiv -\int_{0}^{\infty} ds e^{\nu s \triangle^*} =\int_{-\infty}^{\infty}
d\eta g(\xi,\eta),$$
 where
 $$g(\xi,\eta) \equiv \frac{-1}{\sqrt{2\pi\nu}} {\rm f.p.}\int_{\sqrt{a}}^{\infty}
 \frac{d\sigma}{\sigma^2-a}e^{-\frac{1}{2\nu}\left(\sigma \xi -\eta\sqrt{\sigma^2-a}\right)^2}$$
 and f.p. denotes the finite part of Hadamard, e.g. \cite{Bureau1955, Yosida1956}.
 It can be verified by changing the  variable \textcolor{black}{from $s$ to}
 $\sigma=\sqrt{\frac{a}{1-e^{-2a\tau}}},$ using the solution of the Fokker-Planck  equation
$$e^{\nu \tau \triangle^*}f=\left( \frac{a}{2\pi \nu(1-e^{-2 a \tau})}\right)^{1/2}
 \int_{\mathbb{R}^1}e^{a\tau} f(e^{a\tau} \eta)\exp \left( -\frac{a}{2\nu} \frac{(\xi-\eta)^2}{1-e^{-2 a \tau}}\right) d\eta.$$
As we will consider the 3D Navier-Stokes equations, for which methods of exact solutions are \textcolor{black}{unavailable},
we treat (\ref{ScaledBurgers}) by approximate methods as an illustration. Because the inversion of
$\triangle^*$ is \textcolor{black}{unwieldy}, 
we will seek a workaround by which we can dispense with it.

First we convert it to an integral equation by
the Duhamel principle for the Fokker-Planck operator
$\triangle^*$ 
\begin{align*}
 U(\tau)  = & e^{\nu \tau \triangle^*} U(0)
  -\int_{0}^{\tau} e^{\nu(\tau-s)\triangle^*}\partial\, \frac{U(s)^2}{2}ds &\\
   = & e^{\nu \tau \triangle^*} U(0)
   -\int_{0}^{\tau} e^{\nu s \triangle^*}\partial\,\frac{U(\tau-s)^2}{2}ds. &
\end{align*}
The long-time limit $U_1=\lim_{\tau \to \infty} e^{\nu s \triangle^*} U(0)$ is given by
$$U_1=\left( \frac{a}{2\pi \nu}\right)^{1/2} M e^{-\frac{a}{2\nu}\xi^2}\;\;\mbox{with}
\;\;M=\int_{-\infty}^{\infty} U(0) d\xi.$$
We may consider a number of different iteration schemes. For example, the following option (1), also known as the Picard iteration, requires
the inversion $(\triangle^*)^{-1}$:
$$\mbox{Successive approximation (1):}\;\;U_{n+1} =U_1-\int_{0}^{\infty} e^{\nu s \triangle^*}\partial\,\frac{U_n^2}{2}ds,$$
  $$\mbox{in particular, for}\;\;n=1:\;\;  U_2=U_1-\int_{0}^{\infty} e^{\nu s \triangle^*}\partial\,\frac{U_1^2}{2}ds.$$
\textcolor{black}{Note} that $U_n$ is a steady function at each step.

Alternatively we first consider the steady equation 
$$\triangle^* U\equiv \triangle U +\frac{a}{\nu}(\xi U)_\xi=\frac{1}{\nu}\left( \frac{U^2}{2}\right)_\xi$$
and then introduce iteration schemes:
$$\mbox{Iteration scheme (2a):}\;\;
\triangle U_{n+1} +\frac{a}{\nu}(\xi U_{n+1})_\xi  =\frac{1}{\nu}\left( \frac{U_{n}^2}{2}\right)_\xi,\;\;(n \geq 0)$$
$$\mbox{For}\;\;n=1:\;\;\triangle U_{2} +\frac{a}{\nu}(\xi U_{2})_\xi  =\frac{1}{\nu}\left( \frac{U_{1}^2}{2}\right)_\xi,$$
or
$$\mbox{Iteration scheme (2b):}\;\;\triangle U_{n+1} =-\frac{a}{\nu}(\xi U_{n})_\xi  +\frac{1}{\nu}\left( \frac{U_{n}^2}{2}\right)_\xi,
\;\;(n \geq 1)$$
$$\mbox{For}\;\;n=1:\;\;\triangle U_{2} =\underbrace{-\frac{a}{\nu}(\xi U_{1})_\xi}_{= \triangle U_1}
+\frac{1}{\nu}\left(\frac{U_{1}^2}{2}\right)_\xi.$$
Note that iteration schemes (1) and (2a) coincide with each other at $n=1$.

\subsection{\textcolor{black}{Estimation of the strength of nonlinearity}}
For the Burgers equation we can work out the two kinds of approximations to the second-order
analytically. After \textcolor{black}{straightforward} algebra they are
     \begin{align*}
       \mbox{(1)}\;\; U & \approx  Ce^{-\frac{a}{2\nu}\xi^2}
       \left( 1+\frac{C}{2\nu} \int_0^\xi e^{-\frac{a}{2\nu}\eta^2} d\eta \right),\\
       \mbox{(2b)}\;\; U & \approx  Ce^{-\frac{a}{2\nu}\xi^2}
       +\frac{C^2}{2\nu} \int_0^\xi e^{-\frac{a}{\nu}\eta^2} d\eta,
     \end{align*}
     where $C \approx \sqrt{\frac{a}{2\pi\nu}} M.$
     On this basis we estimate the \textcolor{black}{strength} of the nonlinear term $N(\xi).$
     \textcolor{black}{The source-type solution is a near identity transform of the  Gaussian function.
        In its series expansion in the Reynolds number $Re=M/\nu$ after non-dimensionalisation,
     the nonlinear correction term has $Re$ as its prefactor. }
     \textcolor{black}{Consider the scheme (1), or (2a) equivalently, taking $a/2\nu=1$ without loss of generality. We have
       $$U_2=\frac{M}{\sqrt{\pi}}e^{-\xi^2}\left( 1+ \frac{Re}{2\sqrt{\pi}}\int_0^\xi e^{-\eta^2}d\eta\right).$$
       Separating out the $Re$-dependence, or equivalently assuming that $Re=1,$
       we define $N(\xi)$ by
     $$N(\xi)= \frac{1}{2\sqrt{\pi}}\int_0^\xi e^{-\eta^2}d\eta$$
     so that $U_2 \propto 1+Re N(\xi)$ holds. The strength of nonlinearity is given by
     $$N=\sup_{\xi} N(\xi)=\frac{1}{4}.$$
     The same goes for (2b) from the above expressions for the 1D Burgers equation. Altogether we find
     $$\mbox{(1)}\;\; N= \frac{1}{4}=0.25,$$
     $$\mbox{(2b)}\;N=\frac{1}{4\sqrt{2}}\approx 0.2.$$
  We conclude that the typical \textcolor{black}{strength} of nonlinearity is $N=O(10^{-1})$  irrespective of the choice of schemes.}
  
\subsection{Burgers equations in several dimensions}  

The source-type solution is basically a near-identity function of the Gaussian
form.
It has been seen how the source-type solutions show up in the long-time limit
in one and two spatial dimensions in \cite{Ohkitani2020}. Here we will take a look at
cases in three and higher dimensions.  From the Cole-Hopf transform we have
$$U_i(\bm{\xi},\tau)
=-2\nu\frac{\partial_{\xi_i} \int_{\mathbb{R}^3} \psi_0(\lambda \bm{\eta})
\exp \left(-\dfrac{a}{2\nu}\dfrac{|\bm{\xi}-\bm{\eta}|^2}{1-e^{-2a\tau}}\right)
d\bm{\eta}}{\int_{\mathbb{R}^3} \psi_0(\lambda \bm{\eta})
\exp\left(-\dfrac{a}{2\nu}\dfrac{|\bm{\xi}-\bm{\eta}|^2}{1-e^{-2a\tau}}\right)  
d\bm{\eta}}.$$
\textcolor{black}{As we are going for} the type 2 scale-invariance,  differentiating it twice we find
$$\partial_{\xi_j}\partial_{\xi_k} U_i(\bm{\xi},\tau)
=-2\nu\frac{\partial_{\xi_i} \partial_{\xi_j} \partial_{\xi_k}
  \int_{\mathbb{R}^3} \psi_0(\lambda \bm{\eta})
\exp \left(-\dfrac{a}{2\nu}\dfrac{|\bm{\xi}-\bm{\eta}|^2}{1-e^{-2a\tau}}\right)
d\bm{\eta}}{\int_{\mathbb{R}^3} \psi_0(\lambda \bm{\eta})
\exp\left(-\dfrac{a}{2\nu}\dfrac{|\bm{\xi}-\bm{\eta}|^2}{1-e^{-2a\tau}}\right)  
d\bm{\eta}}+\ldots
$$
$$
=-2\nu\frac{\lambda^3\int_{\mathbb{R}^3} \partial_i \partial_j \partial_k\psi_0(\lambda \bm{\eta})
\exp \left(-\dfrac{a}{2\nu}\dfrac{|\bm{\xi}-\bm{\eta}|^2}{1-e^{-2a\tau}}\right)
d\bm{\eta}}{\int_{\mathbb{R}^3} \psi_0(\lambda \bm{\eta})                                  
\exp\left(-\dfrac{a}{2\nu}\dfrac{|\bm{\xi}-\bm{\eta}|^2}{1-e^{-2a\tau}}\right)  
d\bm{\eta}}+\ldots
$$
The denominator then tends to $K_{ijk}\exp \left(-\frac{a}{2\nu}|\bm{\xi}|^2 \right)$ as $\tau \to \infty,$
where $K_{ijk}=\int_{\mathbb{R}^3} \partial_i \partial_j \partial_k \psi_0( \bm{\eta}) d\bm{\eta},\;
(i=1,2,3).$
Hence
$$\partial_{\xi_j}\partial_{\xi_k} U_i(\bm{\xi},\infty)
=-2\nu \left( \frac{K_{ijk}\exp \left(-\frac{a}{2\nu}|\bm{\xi}|^2 \right)}{F_{ijk}(\bm{\xi})}
+\ldots\right),$$
where the function $F_{ijk}$ is to be determined such that
$\partial_i \partial_j \partial_k F_{ijk} \propto
\exp \left(-\frac{a}{2\nu}|\bm{\xi}|^2 \right),\;\;\textcolor{black}{\mbox{(no summation implied)}}.$
 We can thus take
$$
F_{ijk}(\bm{\xi})=-\frac{K_{ijk}}{2\nu}
\int_0^{\xi_1} \exp \left(-\frac{a \xi^2}{2\nu} \right) d\xi
\int_0^{\xi_2} \exp \left(-\frac{a \eta^2}{2\nu} \right) d\eta
\int_0^{\xi_3} \exp \left(-\frac{a \zeta^2}{2\nu} \right) d\zeta
+1.$$
Therefore \textcolor{black}{after collecting other terms of derivatives} we find in three dimensions, say, with $(i,j,k)=(1,2,3),$
\bel{source_Burgers3D}
\frac{\partial^2 U_1}{\partial \xi_2 \partial \xi_3}
=K_{123}
\exp \left(-\frac{a}{2\nu}(\xi_1^2+\xi_2^2+\xi_3^2)\right)
\frac{1+R(\xi_1, \xi_2,\xi_3)}{(1-R(\xi_1,\xi_2,\xi_3))^3},
\ee
where
$$R(\xi_1, \xi_2,\xi_3)=\frac{K_{123}}{2\nu} 
\int_0^{\xi_1} \exp \left(-\frac{a \xi^2}{2\nu} \right) d\xi
\int_0^{\xi_2} \exp \left(-\frac{a \eta^2}{2\nu} \right) d\eta
\int_0^{\xi_3} \exp \left(-\frac{a \zeta^2}{2\nu} \right) d\zeta$$
denotes the Reynolds number. Because $R$ is small the expression (\ref{source_Burgers3D}) is near-Gaussian.
\textcolor{black}{It can be verified that} $K_{123}=\sqrt{\dfrac{32a^3}{\pi^3 \nu}} \tanh \dfrac{M_{123}}{16\nu},$ where
$M_{123}=\int \frac{\partial^2 U_1}{\partial \xi_2 \partial \xi_3} d\bm{\xi}.$
We can also write
$$\frac{\partial^2 U_1}{\partial \xi_2 \partial \xi_3}
=-2 \nu \frac{\partial^3}{\partial \xi_1 \partial \xi_2 \partial \xi_3} 
\log(1-R(\xi_1, \xi_2,\xi_3)),$$
which reflects the Cole-Hopf transform  more directly, that is, 
$\bm{U}=\nabla_{\bm{\xi}} \phi,\;\phi=-2\nu \log(1-R(\bm{\xi})).$
See {\bf Appendix A} for the general form in $n$-dimensions.

\section{2D Navier-Stokes equation}
\textcolor{black}{
  {\it We briefly recall the self-similar solution of the 2D Navier-Stokes equations,
    where the so-called Burgers vortex appears after dynamic scaling.}}
\subsection{Critical scale-invariance}
The Burgers vortex was  originally introduced to represent the reaction of a vortex under
the influence of the collective effect of surrounding vortices in the ambient medium.
When we write the steady solution in velocity and vorticity
using cylindrical coordinates
$$\bm{u}=(u_r, u_\theta, u_\phi)=(-a r, v(r), 2 a z),\;
\bm{\omega}=(0, 0, \omega(r)),$$
the solution takes the following forms
$$\omega(r)=\frac{a\Gamma}{2\pi\nu}\exp\left(-\frac{a r^2}{2\nu} \right),$$
$$v(r)=\frac{\Gamma}{2\pi r}\left(1-\exp \left(-\frac{a r^2}{2\nu}\right)\right),$$
where
$\Gamma \equiv \int_{\mathbb{R}^2}\omega_0(\bm{x})d \bm{x}$
denotes the velocity circulation.

\textcolor{black}{In two dimensions dynamic scaling transforms take the following form
$$                                                                                        
\bm{\xi}=\frac{\bm{x}}{\sqrt{2 at}},\;\;                                            
\tau=\frac{1}{2 a}\log t,
\bm{\psi}(\bm{x},t)=\bm{\Psi}(\bm{\xi},\tau),$$
$$
\bm{u}(\bm{x},t)=\frac{1}{\sqrt{2at}}\bm{U}(\bm{\xi},\tau),
\omega(\bm{x},t)=\frac{1}{2at}\Omega(\bm{\xi},\tau).
$$
In the 2-dimensional case, in order to achieve the critical scale-invariance of type 2, we must choose
the second spatial derivative of the stream function, which is the vorticity, as the unknown.}

The scaled form of the vorticity equation in two dimensions reads
$$
\dfrac{\partial \Omega}{\partial \tau}+ \bm{U}\cdot \nabla  \Omega
=\nu \triangle \Omega + a \nabla \cdot (\bm{\xi}\Omega),$$
where $\Omega$ satisfies the type 2 scale-invariance.
It is known that the  self-similar solution under scaling has
a mathematically identical form as the Burgers vortex above.
Indeed in the scaled variables the above expression can be written
$$\Omega(\xi)=\frac{a \Gamma}{2\pi\nu}\exp\left(-\frac{a|\bm{\xi}|^2}{2\nu} \right),
\,\bm{\xi}=\frac{\bm{x}}{\sqrt{2at}}.$$

Note that 
$\frac{1}{2at}\Omega(\xi)
=\frac{\Gamma}{4\pi\nu t}\exp\left(-\frac{|\bm{x}|^2}{4\nu t} \right)$
is an exact self-similar decaying solution\footnote{\textcolor{black}{When $a=0$ an exact decaying solution is obtained
by formally replacing $a\to\frac{1}{2t},$ which is known as the Lamb-Oseen vortex.}}
with the following property
$$\lim_{t \to 0} \Omega(\cdot)=\Gamma 
\delta(\bm{x}).$$
It also satisfies the following asymptotic property, for $\omega(\cdot,0) \in L^1,$ 
$$t^{1-\frac{1}{p}}\left\|\omega(\bm{x},t)-\frac{1}{2at}\Omega(\bm{\xi})
\right\|_{L^p} \to 0\;\;\mbox{as}\;\; t \to \infty,$$
where $1 \leq p \leq \infty,$ see e.g.  \cite{GW2005}.
\textcolor{black}{
\subsection{Interpretation}
Because the source-type solution coincides with the linearised solution,
the inhomogeneous terms on the right-hand side of the approximations at each order vanish identically.
Hence there is no way to set up  successive approximations that can capture non-zero
nonlinear corrections. The strength of nonlinearity is identically zero; $N=0$.
}

\section{3D Navier-Stokes equations}
\textit{
We will describe two approaches for handling  the scaled 3D Navier-Stokes equations perturbatively.
First we describe a general framework based on the Green's function.
Second we describe  an iterative approach which is specifically suited for calculations
associated with the 3D Navier-Stokes problem.}

\subsection{\textcolor{black}{Critical scale-invariance}}
We consider the 3D Navier-Stokes equations written in four different dependent
variables. Starting from the vector potential and taking a curl successively
$\bm{u}=\nabla \times \bm{\psi},\; \bm{\omega}=\nabla \times \bm{u},\;
\bm{\chi}=\nabla \times \bm{\omega},$ we have
\bel{NS3D}
\left\{
\begin{array}{l}
\dfrac{\partial \bm{\psi}}{\partial t}
=\dfrac{3}{4\pi}{\rm p.v.}\displaystyle{\int_{\mathbb{R}^3}}
\dfrac{\bm{r} \times( \nabla \times \bm{\psi} (\bm{y}))\,
\bm{r} \cdot (\nabla \times \bm{\psi}  (\bm{y}))}
      {|\bm{r}|^5}\;{\rm d}\bm{y}+\nu\triangle \bm{\psi},\\
\noalign{\vskip 0.2cm}      
\dfrac{\partial \bm{u}}{\partial t}
+ \bm{u}\cdot \nabla  \bm{u} =-\nabla p + \nu\triangle\bm{u},\\
\noalign{\vskip 0.2cm}      
\dfrac{\partial \bm{\omega}}{\partial t}+ \bm{u}\cdot \nabla  \bm{\omega}
=\bm{\omega} \cdot \nabla \bm{u} + \nu \triangle \bm{\omega},\\
\noalign{\vskip 0.2cm}      
\textcolor{black}{
\dfrac{\partial \bm{\chi}}{\partial t}
=\triangle( \bm{u} \cdot \nabla \bm{u}+ \nabla p)  +\nu \triangle  \bm{\chi}
}
\end{array} \right.
\ee
where $\bm{r}\equiv\bm{x}-\bm{y}$ and p.v. denotes a principal-value integral. 
We also have
$ \bm{\omega}=- \triangle \bm{\psi}, \bm{\chi}=-\triangle \bm{u},$
because of the incompressibility condition. The derivation of (\ref{NS3D})$_1$ can be found in \cite{Ohkitani2015}.
\textcolor{black}{
The final fourth equation (\ref{NS3D})$_4$ is obtained by taking the Laplacian of the velocity equations
(\ref{NS3D})$_2$.
For the $\bm{\chi}$ equations we may alternatively take a curl on the vorticity equations to obtain
a form of equation different from the final line in (\ref{NS3D}), which is useful for handling inviscid fluids
(details to be found in {\bf Appendix B}). }
\textcolor{black}{Under dynamic scaling
$$                                                                                        
\bm{\xi}=\frac{\bm{x}}{\sqrt{2 at}},\;\;                                            
\tau=\frac{1}{2 a}\log t,
\bm{\psi}(\bm{x},t)=\bm{\Psi}(\bm{\xi},\tau),$$
$$
\bm{u}(\bm{x},t)=\frac{1}{(2at)^{1/2}}\bm{U}(\bm{\xi},\tau),
\bm{\omega}(\bm{x},t)=\frac{1}{2at}\bm{\Omega}(\bm{\xi},\tau),\,\mbox{and}\,
\bm{\chi}(\bm{x},t)=\frac{1}{(2at)^{3/2}}\bm{X}(\bm{\xi},\tau),    
$$
the 3D Navier-Stokes equations in four different unknowns are transformed respectively to}
\bel{NS3Dscaled}
\left\{
\begin{array}{l}
\dfrac{\partial \bm{\Psi}}{\partial \tau}
=\dfrac{3}{4\pi} {\rm p.v.}\displaystyle{\int_{\mathbb{R}^3}}
\dfrac{\bm{\rho} \times( \nabla \times \bm{\Psi} (\bm{\eta}))\,
\bm{\rho} \cdot (\nabla \times \bm{\Psi}  (\bm{\eta}))}
{|\bm{\rho}|^5}\;{\rm d}\bm{\eta}+\nu\triangle \bm{\Psi} 
+ a(\bm{\xi}\cdot\nabla)\Psi,\\
\noalign{\vskip 0.2cm}      
\dfrac{\partial \bm{U}}{\partial \tau}
+ \bm{U}\cdot \nabla  \bm{U} =-\nabla P + \nu\triangle\bm{U}
+ a(\bm{\xi}\cdot\nabla)\bm{U}
+a \bm{U},\\
\noalign{\vskip 0.2cm}      
\dfrac{\partial \bm{\Omega}}{\partial \tau}+ \bm{U}\cdot \nabla  \bm{\Omega}
=\bm{\Omega} \cdot \nabla \bm{U} + \nu \triangle \bm{\Omega}
+a(\bm{\xi}\cdot\nabla)\bm{\Omega} +2a \bm{\Omega},\\
\noalign{\vskip 0.2cm}
\textcolor{black}{
\dfrac{\partial \bm{X}}{\partial \tau}
=\triangle \left(\bm{U}\cdot\nabla\bm{U}+\nabla P \right)
+\nu \triangle  \bm{X}+a\nabla\cdot(\bm{\xi}\otimes \bm{X}),
}
\end{array}\right.
\ee
where $\bm{\rho}\equiv\bm{\xi}-\bm{\eta}.$

It is to be noted that the coefficient of the linear term increases in number with the increasing order of derivatives
and for the variable $\bm{\chi}$ a divergence is completed in the convective term. 
Observe that type 1 scale-invariance is achieved with $\bm{\Psi}$ and type 2 scale-invariance with  $\bm{X}$.
\subsection{Successive approximations}
Using the Duhamel principle we convert the scaled  Navier-Stokes equations (\ref{NS3Dscaled})$_4$
into integral equations
$$
\bm{X}(\bm{\xi}, \tau)=e^{\nu \tau \triangle^*} \bm{X}_0(\bm{\xi})
+\int_0^{\tau} e^{\nu s \triangle^*}\triangle \left(\bm{U}\cdot\nabla\bm{U}+\nabla P \right)(\bm{\xi},\tau-s)ds.
$$
Here $\triangle^* \equiv \triangle +\frac{a}{\nu}\nabla\cdot(\bm{\xi}\otimes \cdot)$ and
the action of whose exponential operator is given by
\bel{FP3D}
\exp(\nu \tau \triangle^*)f(\cdot)
=\left( \frac{a}{2\pi \nu(1-e^{-2 a \tau})}\right)^{3/2}
\int_{\mathbb{R}^3} e^{3a\tau}f(e^{a\tau}\bm{y})
\exp \left( -\frac{a}{2\nu} \frac{|\bm{\xi}-\bm{y}|^2}{1-e^{-2 a \tau}}\right) d\bm{y}
\ee
for any function $f,$  \textcolor{black}{as can be verified by combining the heat kernel and dynamic scaling transforms.}

The inverse operator associated with
the fundamental solution to the Fokker-Planck equation in 3D is defined by
$$(\nu \triangle^*)^{-1} \equiv -\int_{0}^{\infty} ds e^{\nu s \triangle^*} =\int d\textcolor{black}{\bm{\eta}} g(\bm{\xi},\bm{\eta}),$$
 where \textcolor{black}{the Green's function is given by}
 $$g(\bm{\xi},\bm{\eta}) \equiv \frac{-1}{(2\pi\nu)^{3/2}} {\rm f.p.}\int_{\sqrt{a}}^{\infty}
  \frac{\sigma^2 d\sigma}{\sigma^2-a}e^{-\frac{1}{2\nu}|\sigma \bm{\xi} -\bm{\eta}\sqrt{\sigma^2-a}|^2}.$$
  This can be verified by changing the variable \textcolor{black}{from $s$ to}  $\sigma=\sqrt{\frac{a}{1-e^{-2a\tau}}}$
  in the solution  (\ref{FP3D}) to  the Fokker-Planck  equation.

We consider the steady solution $\bm{X}(\bm{\xi})$
in the long-time limit of $\tau \to \infty$

$$\bm{X}(\bm{\xi})=\bm{X}_1(\bm{\xi}) + \lim_{\tau \to \infty}
\int_0^{\tau} e^{\nu s \triangle^*}\triangle \left(\bm{U}\cdot\nabla\bm{U}+\nabla P \right)(\bm{\xi},\tau-s)ds,
$$
where $\bm{X}_1=\mathbb{P}\bm{M} G$ denotes the leading-order approximation (to be made more explicit in next subsection).
\textcolor{black}{This is one form of integral equations we are supposed to handle.}

On the other hand, steady equations are obtained by assuming $\partial/\partial\tau=0$ in (\ref{NS3Dscaled})$_4$
$$
\triangle^* \bm{X}\equiv \triangle \bm{X}
+\frac{a}{\nu} \nabla\cdot(\bm{\xi}\otimes \bm{X})=-\frac{1}{\nu}
\triangle \left(\bm{U}\cdot\nabla\bm{U}+\nabla P \right),
$$
or
$$
\bm{X}=-\frac{1}{\nu}\left(\bm{U}\cdot\nabla\bm{U}+\nabla P \right)
-\frac{a}{\nu}\triangle^{-1} \nabla\cdot(\bm{\xi}\otimes \bm{X}).
$$
This is yet another  form of the steady Navier-Stokes equations after dynamic scaling.
\textit{It is noted that one of the potential problems associated with the nonlinear
term has been eliminated without having a recourse to the Green's function.}
It is this virtually trivial fact that allows us to set up a simple successive
approximation.

To summarise, the steady Navier-Stokes equations after dynamic scaling can be written as
\be
\bm{X}=-\frac{1}{\nu}\mathbb{P} \left(\bm{U}\cdot\nabla\bm{U}\right)
-\frac{a}{\nu}\triangle^{-1} \nabla\cdot(\bm{\xi}\otimes \bm{X}),
\ee
or, by $\bm{X}=-\triangle \bm{U},$ we can express it solely in terms of $\bm{X}$ as
\bel{scaled.Chi.eq}
\bm{X}=-\frac{1}{\nu}\mathbb{P} \left(\triangle^{-1}\bm{X}\cdot\nabla \triangle^{-1}\bm{X}\right)
-\frac{a}{\nu}\triangle^{-1} \nabla\cdot(\bm{\xi}\otimes \bm{X}).
\ee
This is the set of equations that we need to solve.

In passing we note the following facts before proceeding to the specific results. By the definition of
scaled variables it is easily seen that for $p \ge 1$
    $$t^{\frac{3}{2}\left(1-\frac{1}{p}\right)}
\left\| \bm{\chi}(\bm{x},t)-\frac{\bm{X}(\bm{\xi})}{(2at)^{3/2}}\right\|_{L^p}
=\left\| \bm{X}(\bm{\xi},\tau)-\bm{X}(\bm{\xi}) \right\|_{L^p}.$$
This means that if 
$\left\| \bm{X}(\bm{\xi},\tau)-\bm{X}(\bm{\xi}) \right\|_{L^p} \to 0$ as $\tau \to \infty,$
we have
 $$t^{\frac{3}{2}\left(1-\frac{1}{p}\right)}
\left\| \bm{\chi}(\bm{x},t)-\frac{\bm{X}(\bm{\xi})}{(2at)^{3/2}}
\right\|_{L^p} \to 0\;\;\mbox{as}\;\;  t \to \infty.$$
That is about the long-time asymptotics. On the other hand as time-zero asymptotics
we have
$$\frac{\bm{X}(\bm{\xi})}{(2at)^{3/2}} \to \mathbb{P} \bm{M}\delta
\;\;\mbox{as}\;\;  t \to 0,$$
where $\delta(\cdot)$ is the Dirac mass.

\subsection{Leading-order approximations}
Before discussing the second-order approximation we derive expressions of the first-order
solutions in several different variables.

We will  derive the basic formulas  by solving the heat equation
$$\frac{\partial \bm{\psi}_1}{\partial t}=\nu \triangle \bm{\psi}_1.$$
The first-order approximation is given by
$$\bm{\psi}_1(\bm{x},t)
=\frac{1}{(4\pi \nu t)^{3/2}}\int_{\mathbb{R}^3} \bm{\psi}_0(\bm{y})
\exp \left( -\frac{|\bm{x}-\bm{y}|^2}{4\nu t}\right) d \bm{y}.$$
After applying the dynamic scaling
$$\bm{\psi}_1(\bm{x},t)=\bm{\Psi}_1(\bm{\xi},\tau),
\bm{\xi}=\frac{\bm{x}}{\sqrt{2 at}},\;                                  
\tau=\frac{1}{2 a}\log t,$$
the linearised equations for the vector potential read
$$\frac{\partial \bm{\Psi}_1}{\partial \tau}=a\bm{\xi}\cdot \nabla \bm{\Psi}_1
+\nu \triangle \bm{\Psi}_1.$$
\textcolor{black}{ After dynamic scaling their} solution is given by 
$$\bm{\Psi}_1(\bm{\xi},\tau)=e^{-3 a \tau}
\left( \frac{a}{2\pi \nu(1-e^{-2 a \tau})}\right)^{3/2}
\int_{\mathbb{R}^3} \bm{\Psi}_0(\bm{\eta})
\exp \left( -\frac{a}{2\nu} \frac{|\bm{\xi}-\bm{\eta}e^{-a \tau}|^2}
{1-e^{-2 a \tau}}\right) d\bm{\eta}$$
$$=\left( \frac{a}{2\pi \nu(1-e^{-2 a \tau})}\right)^{3/2}
\int_{\mathbb{R}^3} \bm{\Psi}_0(e^{a\tau}\bm{y})
\exp \left( -\frac{a}{2\nu} \frac{|\bm{\xi}-\bm{y}|^2}
{1-e^{-2 a \tau}}\right) d\bm{y},$$
where $\bm{\Psi}_1$ denotes the first-order approximation and $\bm{\Psi}_0$ the initial data.     
(The same convention applies to $\bm{X}_1$ and $\bm{X}_0$ in the following.)
Taking a curl with respect to $\bm{\xi}$ three times we find  the expressions for the vorticity curl
$$\bm{X}_1(\bm{\xi},\tau)
=\left( \frac{a}{2\pi \nu(1-e^{-2 a \tau})}\right)^{3/2}
\int_{\mathbb{R}^3}  e^{3a\tau}\bm{X}_0(e^{a\tau}\bm{y})
\exp \left( -\frac{a}{2\nu} \frac{|\bm{\xi}-\bm{y}|^2}
{1-e^{-2 a \tau}}\right) d\bm{y}.$$
 For well-localised initial data we make use of the formula
$\lambda^3 \bm{X}_0(\lambda\bm{y}) \to \bm{M}\delta(\bm{y})$ where $\bm{M}=\int \bm{X}_0 d\bm{y}.$
Noting that $\mathbb{P}\bm{X}_0=\bm{X}_0$ we have\footnote{If the initial condition satisfies the similarity condition
$\lambda^3 \bm{X}_0(\lambda\bm{y})=\bm{X}_0(\bm{y})\;\mbox{for}\,\forall \lambda (>0),$ we have
  $\bm{X}_1(\bm{\xi},\tau)\to
\left( \frac{a}{2\pi \nu}\right)^{3/2}
\int_{\mathbb{R}^3} \bm{X}_0(\bm{y})
\exp \left( -\frac{a}{2\nu} |\bm{\xi}-\bm{y}|^2 \right) d\bm{y}
= \bm{X}_0*G.$ In this case $\bm{X}_0(\bm{\xi})$ is singular like $\sim |\bm{\xi}|^{-3}$.}

\begin{eqnarray}
\bm{X}_1(\bm{\xi},\tau)
&=&\left( \frac{a}{2\pi \nu(1-e^{-2 a \tau})}\right)^{3/2}
\int_{\mathbb{R}^3} e^{3a\tau}(\mathbb{P}\bm{X}_0(e^{a\tau}\bm{y}))
\exp \left( -\frac{a}{2\nu} \frac{|\bm{\xi}-\bm{y}|^2}{1-e^{-2 a \tau}}\right) d\bm{y} \nonumber\\
&=&\left( \frac{a}{2\pi \nu(1-e^{-2 a \tau})}\right)^{3/2}
\int_{\mathbb{R}^3} e^{3a\tau}\bm{X}_0(e^{a\tau}\bm{y})
\mathbb{P}\exp \left( -\frac{a}{2\nu} \frac{|\bm{\xi}-\bm{y}|^2}{1-e^{-2 a \tau}}\right) d\bm{y} \nonumber\\
&\to& \mathbb{P}\bm{M}G\:\:\mbox{as}\;\; \tau \to \infty, \nonumber
\end{eqnarray}
where $G(\bm{\xi})\equiv\left( \frac{a}{2\pi \nu}\right)^{3/2}
\exp \left(-\frac{a}{2\nu}|\bm{\xi}|^2 \right)$ and $\bm{M}=\int \bm{X}_0 d\bm{\xi}.$
This is the leading-order approximation for the scaled 3D Navier-Stokes equations.

The first-order (that is, the leading-order) approximation obtained above can be calculated explicitly because the Gaussian
function is a radial function.
\textcolor{black}{Care should be taken that the leading-order
 approximations themselves are \textit{not} radial because of the incompressibility condition.}
Indeed, in terms of the vorticity curl the first-order approximation is given by
\textcolor{black}{
  \begin{eqnarray}
X_i &=& M_j\left( \frac{a}{2\pi\nu}\right)^{3/2}
\left(\delta_{ij}-\partial_i \partial_j \triangle^{-1} \right)
\exp\left(-\frac{a}{2\nu}r^2\right),\;\; (i=1,2,3,) \\
 &=&M_i\left(\frac{\mu}{\pi}\right)^{3/2}\exp(-\mu r^2)
+M_j\partial_j\partial_i \frac{{\rm erf}(\sqrt{\mu}r)}{4 \pi r}\nonumber \\
&=&M_j \left(\delta_{ij}-\frac{\xi_i \xi_j}{r^2} \right)
\left(\frac{\mu}{\pi}\right)^{3/2} e^{-\mu r^2}
-M_j \left(\frac{\delta_{ij}}{r^3}-\frac{3\xi_i \xi_j}{r^5}
\right) \left\{  \frac{{\rm erf}(\sqrt{\mu}r)}{4 \pi} \textcolor{black}{-}\frac{r}{2\mu}
\left(\frac{\mu}{\pi}\right)^{3/2} e^{-\mu r^2} \right\},\nonumber
\noindent \end{eqnarray}}
\noindent where $\mu=a/(2\nu),$ $r=|\bm{\xi}|,$ $M_j=\int X_j d\bm{\xi},\;(j=1,2,3)$
and  \textcolor{black}{summation is implied on repeated indices.
In the second line we computed $\triangle^{-1}$ for the Gaussian function using
$\triangle=\frac{1}{r^2}\frac{d}{dr} \left(r^2 \frac{d}{dr}\right)$  and the final line by direct computations.
Clearly the final expression is \textit{not} radial.}

Hereafter in this subsection we take \textcolor{black}{$\mu=1$} for simplicity.
Note that $\triangle^{-n} e^{-r^2}$ for $n=1,2,3$ can be evaluated by quadratures
and their explicit form are as follows, which can be obtained  most conveniently with
the assistance of computer algebra. The results are
\begin{eqnarray}
\triangle^{-1} e^{-r^2}&=&-\frac{1}{2r}\int_0^r e^{-s^2}ds
=-\frac{\sqrt{\pi}}{4r}{\rm erf}(r),\\
\triangle^{-2} e^{-r^2}&=&-\frac{1}{2r}
\int_0^r ds \int_0^s ds' \int_0^{s'} e^{-s''^2}ds'' \nonumber\\
&=&-\frac{e^{-r^2}}{8}
-\frac{\sqrt{\pi}}{8}\left(r+\frac{1}{2r}\right){\rm erf}(r),\\
\triangle^{-3} e^{-r^2}&=&-\frac{1}{2r}
\int_0^r ds \int_0^s ds' \int_0^{s'} ds'' \int_0^{s''} ds''' 
\int_0^{s'''} ds''''  e^{-s''''^2},\nonumber\\
&=&-\frac{1}{384r}
\left[(4r^3+10r)e^{-r^2}+4\sqrt{\pi}\left(r^4+3r^2+\frac{3}{4}\right)
{\rm erf}(r) \right].
\end{eqnarray}

With these at hand the expressions for the \textcolor{black}{several} different unknowns,
including the vorticity curl above, are ($i=1,2,3$)
\begin{eqnarray}
B_i&=&\frac{1}{\pi^{3/2}}
\left(M_i  \triangle^{-2} e^{-r^2} 
- M_j \partial_i \partial_j \triangle^{-3} e^{-r^2} \right),\label{B}\\
\textcolor{black}{\Psi_i}&=&\frac{\epsilon_{ijk}M_j \textcolor{black}{\xi_k}}{8\pi^{3/2}} \textcolor{black}{J(r)},\label{Psi}\\
U_i&=&-\frac{1}{\pi^{3/2}} \left(M_i  \triangle^{-1} e^{-r^2} 
- M_j \partial_i \partial_j \triangle^{-2} e^{-r^2} \right),\label{U}\\
\Omega_i&=& \frac{\epsilon_{ijk}M_j\textcolor{black}{\xi_k}} {4\pi^{3/2}} \textcolor{black}{H(r)}, \label{Omega}\\
X_i&=&\frac{1}{\pi^{3/2}}
\left(M_ie^{-r^2} - M_j \partial_i \partial_j \triangle^{-1} e^{-r^2} \right), \label{Chi}
\end{eqnarray}
where
$\textcolor{black}{\bm{\Psi}}=\nabla \times \bm{B}, \bm{U}=\nabla \times \textcolor{black}{\bm{\Psi}},  
\bm{\Omega}=\nabla \times \bm{U},\bm{X}=\nabla \times \bm{\Omega}$
and ${\rm erf}(r)\equiv\frac{2}{\sqrt{\pi}}\int_0^r e^{-t^2}dt$ denotes the error function.
\textcolor{black}{Here for convenience  we have introduced two functions 
  $$H(r)\equiv\frac{\sqrt{\pi}{\rm erf}(r)-2re^{-r^2}}{r^3},\;
 J(r)\equiv\frac{re^{-r^2}+\sqrt{\pi}{\rm erf}(r)\left(r^2-\frac{1}{2}\right)}{r^3},$$
both of which  are continuous at $r=0$ (See Figure 1 below).}
Because all the fields are incompressible we also have
$\bm{U}=-\triangle \bm{B}, \bm{\Omega}=- \triangle \textcolor{black}{\bm{\Psi}}, \bm{X}=-\triangle \bm{U}.$

\begin{figure}[ht]
\includegraphics[scale=0.4,angle=0]{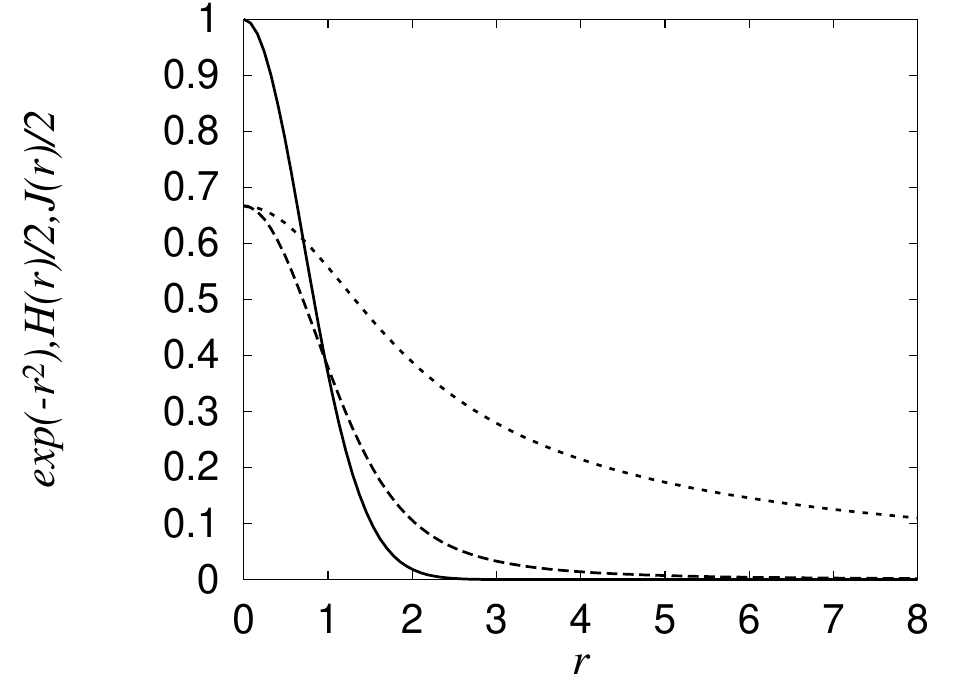}
\caption{Comparison of $\exp(-\xi^2)$ (solid),  $\pi^{3/2}\chi_1(\xi) = H(\xi)/2$ (dashed)
  and $J(\xi)/2$ (dotted). See the text for their definitions. Note that $H(0)=J(0)=2/3.$ }
\label{GvsPG}
\end{figure}

\textcolor{black}{Regarding the derivations of (\ref{B})-(\ref{Chi}),
applying $\triangle^{-1}$ to (\ref{Chi}) repeatedly we get  (\ref{U}) and  (\ref{B}), respectively.
Then, taking a curl of (\ref{B}) we get (\ref{Psi}) and taking a curl of (\ref{U})
we get (\ref{Omega}). Finally simpler forms of (\ref{U}) and (\ref{Chi}) are obtained by
taking a curl of  (\ref{Psi}) and  (\ref{Omega}), respectively. The final results in vectorial form read
\begin{eqnarray}
\bm{\Psi} &=& \frac{\bm{M}\times\bm{\xi}}{8\pi^{3/2}} J(r), \label{vPsi}\\
\bm{U}&=&\frac{\rm{erf}(r)}{4\pi r}\left(\bm{M}-\frac{(\bm{M}\cdot\bm{\xi})\bm{\xi}}{r^2} \right)
-\frac{J(r)}{8\pi^{3/2}}\left(\bm{M}-\frac{3(\bm{M}\cdot\bm{\xi})\bm{\xi}}{r^2} \right),\label{vU}\\
\bm{\Omega} &=& \frac{\bm{M}\times\bm{\xi}} {4\pi^{3/2}}H(r),\label{vOmega}\\
\bm{X}&=&\frac{e^{-r^2}}{\pi^{3/2}}\left(\bm{M}-\frac{(\bm{M}\cdot\bm{\xi})\bm{\xi}}{r^2} \right)
-\frac{H(r)}{4\pi^{3/2}}\left(\bm{M}-\frac{3(\bm{M}\cdot\bm{\xi})\bm{\xi}}{r^2} \right). \label{vChi}
\end{eqnarray}
}
It is of interest to observe that $\bm{U}\cdot\bm{\Omega}=\bm{\Omega}\cdot\bm{X}=0$.
Using the above formula (\ref{Omega}) with $\bm{M}=(1,1,1)$ it is instructive to compare a component of
the Gaussian function $\frac{1}{\pi^{3/2}}e^{-\textcolor{black}{\xi}^2}$ with that of
the vorticity curl
\bel{Leading2}
X_1(\textcolor{black}{\xi},0,0)=\frac{\sqrt{\pi}{\rm erf}(\textcolor{black}{\xi})
  -2xe^{-\textcolor{black}{\xi}^2}}{2\pi^{3/2}\textcolor{black}{\xi}^3}\;\;\;\;
\textcolor{black}{\left(=\frac{H(\xi)}{2\pi^{3/2}} \right)}.
\ee
Figure \ref{GvsPG} shows how $X_1(\xi,0,0)$ is affected by the incompressible condition (solenoidality),
in particular the peak value at $\xi=0$ is reduced by a factor of $2/3$.

\subsection{\textcolor{black}{Estimation of the strength of nonlinearity}}
\textcolor{black}{
We first describe the numerical methods employed to obtain the approximate solutions.
We use a centered finite-difference scheme in a box  $[-L,L]^3$ of size $L,$ with
discretised coordinates
$(x_i,y_j,z_k)=(i\Delta h, j\Delta h,k\Delta h)$ where $i,j,k=-N,\ldots,N$
and $\Delta h=L/N$.}

\textcolor{black}{An important step in the determination of the second-order
approximation is the inversion of the Laplacian operator. This was done by using a
Poisson solver in the Intel MKL library. Parameters used are $L=40, N=800,$ hence
$\Delta h=0.05$.
The box size $L$ was chosen large enough to capture the decay of
$\bm{X}(\bm{\xi})$ near the boundaries and $N$ was chosen large enough to resolve spatial
structure in the center of the domain.
As a code validation we calculated (4.6) numerically and compared against
the analytical expression (\ref{Leading2}) and confirmed their agreement
(figure omitted).}

\textcolor{black}{
To evaluate the perturbation we make use of the iteration scheme (2b) illustrated in the previous
section for simplicity.
Consider a series expansion of the source-solution in $Re$. In comparison to the leading-order the  nonlinear correction is proportional
to $Re$.\footnote{\textcolor{black}{After full non-dimensionalisation $\widehat{\bm{X}}=\dfrac{\nu\bm{\chi}(\bm{x},t)}{(2at)^{3/2}}$ we have
  $\widehat{\bm{X}}_2=\mathbb{P}\widehat{\bm{M}}G                                                                                     
  -\mathbb{P} \left(\triangle^{-1} \mathbb{P}\widehat{\bm{M}}G\cdot\nabla \triangle^{-1} \mathbb{P}\widehat{\bm{M}}G\right),$
  where $\widehat{\bm{M}}=M/\nu.$ As $|\widehat{\bm{M}}|=Re$ it is clear that the nonlinear term has a factor of $Re$ in comparison to the
  leading-order approximation.}}  Separating out $Re,$ or equivalently assuming that $Re=1,$
we define  the strength of nonlinearity by the remaining factor.}
\textcolor{black}{
To be more specific the second-order solution in this case is given  by
  \begin{eqnarray}
  \bm{X}_2&=&\bm{X}_1-\frac{1}{\nu}\mathbb{P} \left(\triangle^{-1}\bm{X}_1\cdot\nabla \triangle^{-1}\bm{X}_1\right) \nonumber\\
 &=&\mathbb{P}\bm{M}G -\frac{1}{\nu}\mathbb{P} \left(\triangle^{-1}\mathbb{P}\bm{M}G \cdot\nabla \triangle^{-1}\mathbb{P}\bm{M}G \right) \nonumber\\
 &=&\mathbb{P}\bm{M}G -Re\,\mathbb{P} \left(\triangle^{-1}\mathbb{P}\bm{M}G \cdot\nabla \triangle^{-1}\mathbb{P}\bm{M}G \right)/|\bm{M}|. \nonumber
 \end{eqnarray}
  Separating out the $Re$-dependence we define $N(\bm{\xi})$ by
  \bel{N}
  N(\bm{\xi})=\frac{\left|\mathbb{P} \left(\triangle^{-1}\mathbb{P}\bm{M}G \cdot\nabla \triangle^{-1}\mathbb{P}\bm{M}G \right)\right|/|\bm{M}|}
  {\sup_{\bm{\xi}} |\mathbb{P}\bm{M}G|}
\ee
and put $N=\sup_{\bm{\xi}}N(\bm{\xi})$.}
\begin{figure}[ht]
\begin{minipage}{0.5\linewidth}
\includegraphics[scale=0.4,angle=0]{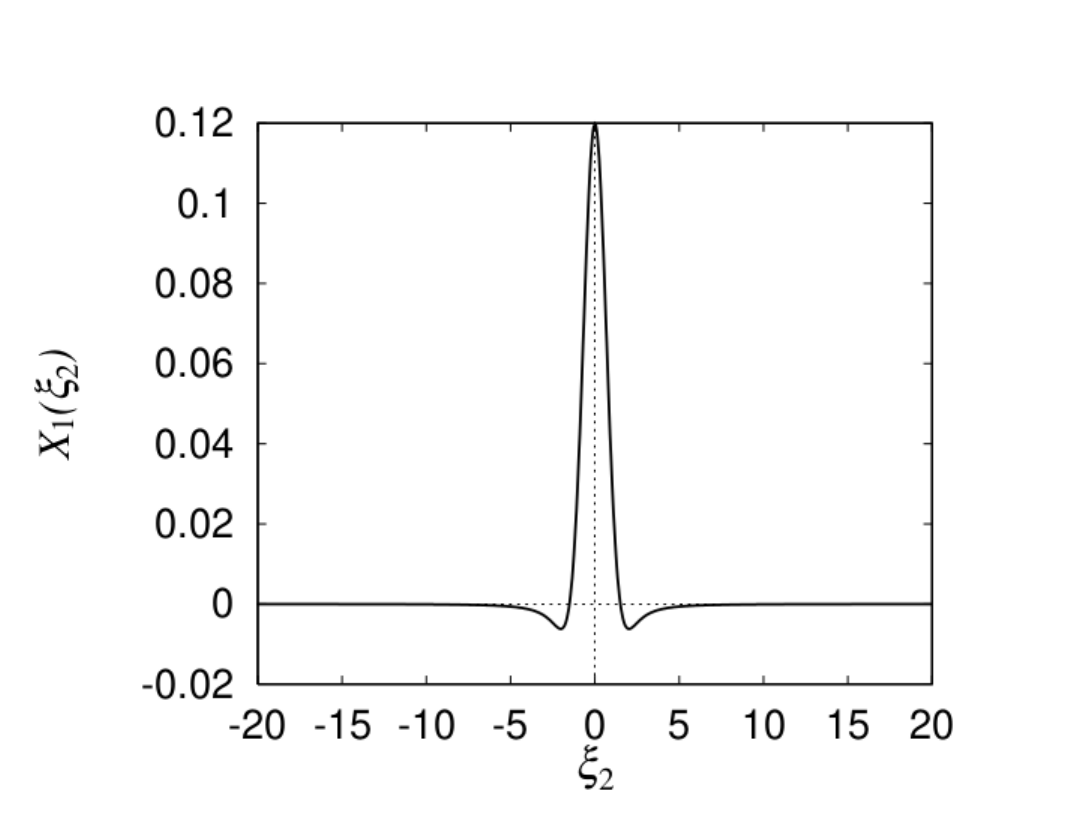}
\caption{The first order term, i.e. the $\xi_1$-component of $\bm{X}_1=\mathbb{P}\bm{M} G$ as a function of $\xi_2$.}
\label{NS1st-order}
\end{minipage}
\begin{minipage}{0.5\linewidth}
\includegraphics[scale=0.4,angle=0]{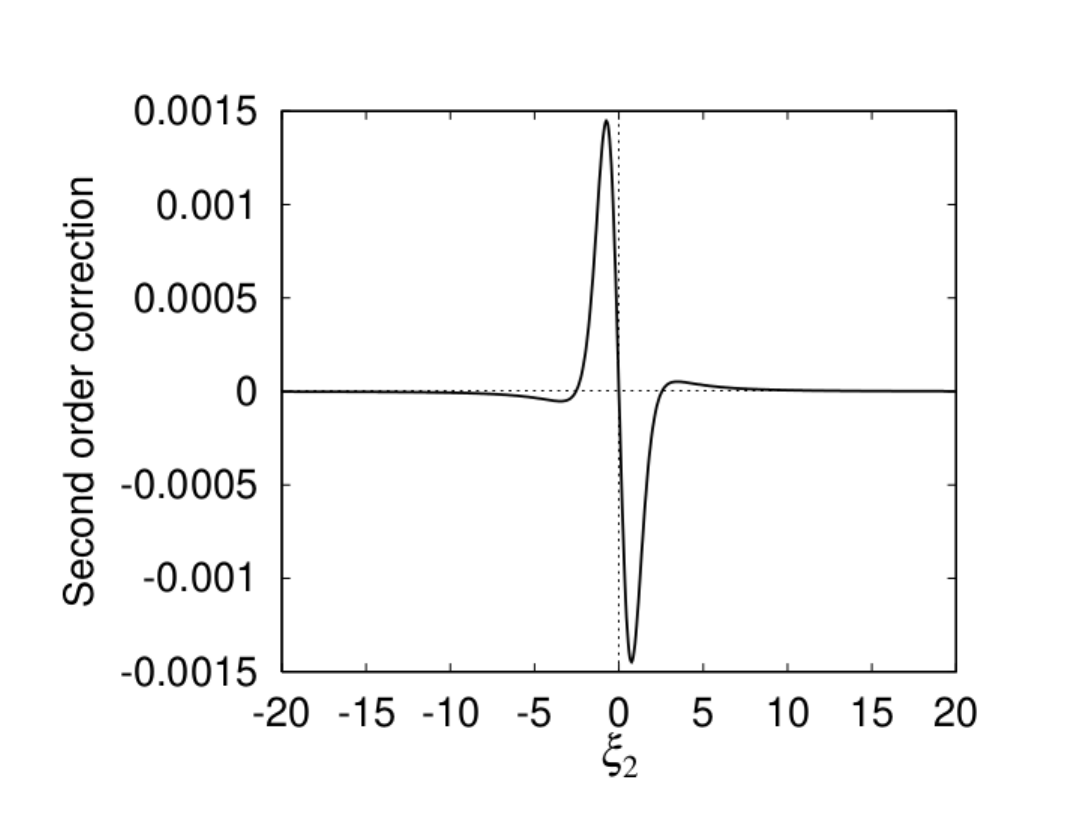}
\caption{The $\xi_1$-component of the second order correction, i.e. the nonlinear term
  $\frac{1}{\nu}\mathbb{P}\left(\triangle^{-1}\bm{X}_1\cdot\nabla \triangle^{-1}\bm{X}_1\right)$
 depicted as in Fig.2.}
\label{NS2nd-order}
\end{minipage}
\end{figure}

It turns out that the $\xi_1$-component of the second-order correction \textcolor{black}{along the line $(\xi_1,0,0)$} is identically equal to zero
\textcolor{black}{owing to the radial symmetry of the Gaussian function.}
\textcolor{black}{For this reason we show} in Figure \ref{NS1st-order}
the \textcolor{black}{$\xi_1$}-component of the first-order approximation $\bm{X}_1$ as a function
of $\xi_2$ along the line  \textcolor{black}{$(0,\xi_2,0)$}. It has a peak at the origin whose height is approximately 0.12.
Accordingly we show in Figure \ref{NS2nd-order} the \textcolor{black}{$\xi_1$-}component of the second-order correction  due to the nonlinearity
$\frac{1}{\nu}\mathbb{P} \left(\triangle^{-1}\bm{X}_1\cdot\nabla \triangle^{-1}\bm{X}_1\right)$
as a function of \textcolor{black}{$\xi_2$}.
It has  double peaks near the origin, but their value is small and is about 0.0015.
Noting $Re=\frac{M}{\nu}=\frac{1}{1/2}=2,$ by (\ref{N})
we can estimate the \textcolor{black}{strength} of nonlinearity in that cross-section as
$$ N \approx \frac{0.0015}{2 \times 0.12} \approx 6 \times 10^{-3}.$$
Actually the maximum value of the nonlinear term in the above sense
in $\mathbb{R}^3$ is 0.0022, not much different from the above value.
Thus the strength of nonlinearity is at most $N \approx 9\times 10^{-3}$ and we conclude
$$N=O(10^{-2})\;\;\mbox{for the 3D Navier-Stokes equations.}$$
It should be noted that it is much smaller than the value of $N$ for the Burgers equations, whose solutions are known to
remain regular all time. Because the difference between the  Navier-Stokes  and  Burgers equations is the presence or absence 
of the incompressibility condition,  it is the incompressibility that makes the value of $N$ smaller for the  Navier-Stokes
equations. On a practical side this also means that even if we add the second-order correction
to the first-order term at low Reynolds number, say $Re=1,$  the superposed solution is virtually indistinguishable from
the first-order approximation. \textcolor{black}{In the final period of decay the Navier-Stokes flows are very close to (\ref{vChi}),
to within $1\%$ at $Re=1,$ which may be regarded as the building blocks for representing the late-stage of evolution, that is, 
as the counterpart of the Burgers vortex in two dimensions.}

\section{Summary and outlook}
We have studied self-similar solutions to the fluid dynamical equations
with particular focus on the so-called source-type solutions.
As an illustration of successive approximation schemes we have discussed the 1D Burgers equation
which is exactly soluble. In this case the velocity is the most convenient choice for its analysis.
We have introduced a method of quantitatively assessing the \textcolor{black}{strength} of nonlinearity $N$ using the source-type solutions.
Similar analyses have been carried out for higher-dimensional Burgers equations.
For the Burgers equations  we find $N=O(0.1)$ in any dimensions.

We then move on to consider the 2D and 3D Navier-Stokes equations.
In two dimensions we review the known results done using the vorticity.
In three dimensions the most convenient choice of the unknown is the vorticity curl.
We have formulated the dynamically-scaled equations using that variable and set up the successive
approximation schemes. We have found that the second-order correction stemming from the nonlinear term
\textcolor{black}{gives rise to} $N \approx 0.01$, an order of magnitude smaller than that for the Burgers equations.
\textcolor{black}{We are led to conclude that the incompressible condition makes $N$ smaller for the Navier-Stokes
equations than for the Burgers equations.}

The current approach relies on perturbative treatments.
It may be challenging, but worthwhile to study the functional form of the solution by non-perturbative
methods for further theoretical developments.
It is also of interest to seek a fully non-linear solution by numerical methods. 
It is noted that this is at least one order of magnitude smaller than $N$ found for the Burgers equations
whose solutions are known to remain regular all the time.
As an application of the source-type solution, it is useful to characterise the late stage of
statistical solutions of the Navier-Stokes equations \cite{Ohkitani2020}.

\appendix
\section{Source solution to the  Burgers equations in $n$ dimensions}
The source-type solution to the $n$-dimensional Burgers equations
takes the following form
$$
\frac{\partial^{n-1}U_1}{\partial \xi_2\ldots\partial \xi_n}=K_{1\ldots n}
\exp\left( -\frac{a}{2\nu}|\bm{\xi}|^2\right) \frac{P_n (-R(\bm{\xi}))}{(1-R(\bm{\xi}))^n},
$$
where  $P_n(s)$'s are polynomials to be constructed below,
\begin{eqnarray}
&R(\bm{\xi})=&\frac{K_{1\ldots n}}{2\nu} \Pi_{j=1}^{n}
\int_{0}^{\xi_j}\exp\left( -\frac{a}{2\nu}\xi^2\right) d\xi, \nonumber\\
&K_{1\ldots n}=&2\nu\left(\frac{2a}{\pi \nu}\right)^{n/2} \tanh \frac{M_{1\ldots n}}{2^{n+1}\nu} \approx
\left(\frac{a}{2\pi \nu}\right)^{n/2} M_{1\ldots n}\;\;(\mbox{for}\;\;M_{1\ldots n}/\nu \ll 1)\nonumber
\end{eqnarray}
and $M_{1\ldots n}=\int \dfrac{\partial^{n-1}U_1}{\partial \xi_2\ldots\partial \xi_n}
d\bm{\xi}.$

{\bf Construction}\\
Consider a function $s(x_1,x_2,\ldots,x_n)$ separable in $n$ variables
$x_1, x_2, \ldots, x_n$
$$s(x_1,x_2,\ldots,x_n)=F(x_1) F(x_2)\ldots F(x_n),$$
where $F(\cdot)$ is a smooth function and $n \in \mathbb{N}$. Define another function
$\phi$ by 
$$\phi(x_1,x_2,\ldots,x_n)=\log (1+s),$$
then we have
$$
\frac{\partial^n \phi}{\partial x_1 \partial x_2 \ldots \partial x_n}
=\frac{\partial^n s}{\partial x_1 \partial x_2 \ldots \partial x_n}
\frac{P_n(s)}{(1+s)^n},
$$
where $P_n(s)$ is a sequence of  polynomials \textcolor{black}{in $s$} of degree \textcolor{black}{not exceeding}  $n-2.$ 
In fact, it is given by
$$P_n(s)=(1+s)^n \left( \frac{d}{ds}s \right)^n \frac{\log(1+s)}{s},
\;\;\mbox{for}\;\;(n=1,2,\ldots)$$
or equivalently,
$$P_n(s)=(1+s)^n \sum_{k=0}^\infty (k+1)^{n-1} (-s)^k.$$ 
The first four of them are $P_1(s)=P_2(s)=1, P_3(s)=1-s, P_4(s)=1-4s+s^2.$

{\bf Proof} (Due to Yuji Okitani.)\\
$$\phi_{x_1}=s_{x_1}\frac{1}{1+s}.$$
Noting $s_{x_2} s_{x_1} = s s_{x_1 x_2},$ we have
$$\phi_{x_1 x_2}= s_{x_1 x_2}\frac{1}{1+s}-s_{x_1}\frac{1}{(1+s)^2} s_{x_2}
=s_{x_1 x_2}\left(\frac{1}{1+s}-\frac{s}{(1+s)^2} \right)=s_{x_1 x_2}\frac{1}{(1+s)^2},$$
while the penultimate expression of which may also be written
$$=s_{x_1 x_2}\underbrace{\left(\frac{1}{1+s}+s\frac{d}{ds}\frac{1}{1+s} \right)}_{\equiv f_2(s)}.$$
Likewise we have
$$\phi_{x_1 x_2 x_3}=s_{x_1 x_2 x_3}
\underbrace{\left(\frac{1}{(1+s)^2}+s\frac{d}{ds}\frac{1}{(1+s)^2} \right)}_{\equiv f_3(s)}.$$
Hence in general the recursion relationship is 
$$f_{n+1}(s)=f_n(s)+s\frac{d}{ds} f_n(s)=\frac{d}{ds}(s f_n(s)),$$
where $f_n=\frac{P_n}{(1+s)^n}.$

Alternatively, by $\phi(s)=\sum_{k=1}^\infty (-1)^{k-1}\dfrac{s^k}{k},$
we compute
$$\phi_{x_1}=\sum_{k=1}^\infty (-1)^{k-1} s^{k-1} s_{x_1},$$
$$\phi_{x_1 x_2}=\sum_{k=1}^\infty (-1)^{k-1}\{(k-1) s^{k-2} s_{x_2} s_{x_1}
+ s^{k-1} s_{x_1 x_2}\}=\sum_{k=1}^\infty (-1)^{k-1} ks^{k-1}  s_{x_1 x_2},$$
and
$$\phi_{x_1 x_2 x_3}=\sum_{k=1}^\infty (-1)^{k-1}\{ k(k-1) s^{k-2}s_{x_3} s_{x_1 x_2}
+ k s^{k-1}  s_{x_1 x_2 x_3} \}
=s_{x_1 x_2 x_3} \sum_{k=1}^\infty (-1)^{k-1} k^2 s^{k-1}.$$
In general we find
$$\phi_{x_1 x_2\ldots  x_n}=s_{x_1 x_2 \ldots x_n}\sum_{k=1}^\infty (-1)^{k-1} k^{n-1} s^{k-1},$$
and hence
$$f_n(s)=\sum_{k=1}^\infty  k^{n-1} (-s)^{k-1}.\;\;\square$$ 

Using the general expression we can estimate the nonlinearity $N$ in $n$-dimensional cases.
Write $P_n(-R)=1+c_n R+\ldots$ for small $R,$ we have
$$\frac{P_n(-R)}{(1-R)^n}\approx(1+c_n R+\ldots)(1+nR+\ldots)\approx 1+(c_n+n)R+\ldots,$$
$$R(\bm{\xi})\approx \left(\frac{a}{2\pi \nu}\right)^{n/2} \frac{M_{1\ldots n}}{2\nu}
\left(\frac{\pi \nu}{2a}\right)^{n/2}\approx \frac{Re}{2^{n+1}},$$
so we find
$$\frac{\partial^{n-1}U_1}{\partial \xi_2\ldots\partial \xi_{n-1}}\approx
\left(\frac{a}{2\pi \nu}\right)^{n/2} M_{1\ldots n} \left(1+\frac{c_n+n}{2^{n+1}}Re+\ldots \right).$$
As $c_n=2^{n-1}-n,$ we deduce $N=\frac{c_n+n}{2^{n+1}}=\frac{1}{4}$ for all $n\;(\ge 1).$
Hence the estimate obtained in the iteration (1) in Section 2(c) holds valid in any dimensions.

\section{Derivation of the vorticity curl equations}
Recalling vector identities
$$\nabla(\bm{A}\cdot\bm{B})=\bm{A}\cdot\nabla \bm{B}+\bm{B}\cdot\nabla \bm{A}
+\bm{A}\times {\rm rot}\bm{B} +\bm{B}\times {\rm rot}\bm{A}$$
$${\rm rot}(\bm{A}\times\bm{B})=-\bm{A}\cdot\nabla \bm{B}+\bm{B}\cdot\nabla \bm{A}
+\bm{A}{\rm div}\bm{B} +\bm{B}{\rm div}\bm{A}$$
and adding them and solving for $\bm{B}\cdot\nabla \bm{A},$ we have
$$\bm{B}\cdot\nabla \bm{A}=\frac{1}{2}\left\{ \nabla (\bm{A}\cdot\bm{B})+ {\rm rot}(\bm{A}\times\bm{B})
-\bm{A}\times {\rm rot}\bm{B}- \bm{A}{\rm div}\bm{B} -\bm{B}\times {\rm rot}\bm{A}+\bm{B}{\rm div}\bm{A}\right\}.$$
Taking $\bm{A}=\bm{\omega}, \bm{B}=\bm{u},$
we find
$$(\bm{u}\cdot\nabla)\bm{\omega}
=\frac{1}{2}\left\{ \nabla (\bm{u}\cdot\bm{\omega})
+\nabla \times (\bm{\omega}\times\bm{u})-\bm{u}\times\bm{\chi}\right\}.$$
We then compute
\begin{eqnarray}
\frac{\partial \bm{\chi}}{\partial t}&=&\nabla \times \left\{\nabla \times (\bm{u} \times \bm{\omega}) \right\} \nonumber\\
&=&\nabla \times (\bm{\omega}\cdot \nabla)\bm{u} -\nabla \times (\bm{u}\cdot \nabla)\bm{\omega}\nonumber\\
&=&\nabla \times (\bm{\omega}\cdot \nabla)\bm{u} -\frac{1}{2}\nabla\times\left\{ \nabla (\bm{u}\cdot\bm{\omega})
+\nabla \times (\bm{\omega}\times\bm{u})-\bm{u}\times\bm{\chi}\right\}  \nonumber\\
&=&\nabla \times (\bm{\omega}\cdot \nabla)\bm{u}+\frac{1}{2} \underline{\nabla\times \nabla \times (\bm{u} \times \bm{\omega})}
+\frac{1}{2}\nabla \times(\bm{u}\times\bm{\chi}), \nonumber
\end{eqnarray}
Identifying the underlined part as the right-hand side, we obtain
$$\frac{\partial \bm{\chi}}{\partial t}=
2 \nabla \times (\bm{\omega}\cdot \nabla)\bm{u}+\nabla \times(\bm{u}\times\bm{\chi}),$$
that is,
$$\frac{D \bm{\chi}}{D t}= \bm{\chi}\cdot \nabla\bm{u}+
2 \nabla \times (\bm{\omega}\cdot \nabla)\bm{u}.$$

\section{Steady Fokker-Planck equation}
\subsection{One-dimensional case}
Because 
\bel{1D_FP}
\frac{\partial^2 U}{\partial \xi^2}
+\frac{a}{\nu}\frac{\partial}{\partial \xi}\left( \xi U \right)=0
\ee
is a second-order equation, its general solution has {\it two} constants of integration.
We solve it paying attention to the boundary conditions. 
First, upon integration we have
$$ \frac{\partial U}{\partial \xi}
+\frac{a}{\nu} \xi U =C,$$
for some constant $C$. If $U$ decays sufficiently rapidly
as $|\xi| \to \infty$, then $C=0$. Otherwise it's possible to have $C \ne 0$.
Let us proceed keeping $C$ and write
$$e^{-\frac{a}{2\nu}\xi^2} \partial_\xi(e^{\frac{a}{2\nu}\xi^2}U(\xi))=C.$$
A further integration gives
\begin{eqnarray}
U(\xi)&=&C' \exp\left(-\frac{a}{2\nu}\xi^2\right)
+C\exp\left(-\frac{a}{2\nu}\xi^2\right)\int_0^\xi \exp\left(\frac{a}{2\nu}\eta^2\right)
d\eta,\nonumber \\
&=& C' \exp\left(-\frac{a}{2\nu}\xi^2\right)
+C \sqrt{\frac{2\nu}{a}}D\left(\sqrt{\frac{a}{2\nu}}\xi\right),\nonumber
\end{eqnarray}
with another constant $C'$.
Here we have assumed that
$|U|, |\partial_\xi U| \to 0$ as $|\xi| \to \infty.$ 
However, if $\xi U(\xi) \to  \mbox{const},$ the second term survives with
$C \ne 0$.
The function $$D(x) \equiv  e^{-x^2}\int_0^x e^{y^2} dy$$
is the Dawson's integral \cite{OLB2010}, which behaves $\approx \frac{1}{2x}$ as
$|x|\to \infty.$  It also satisfies
$$D(x)=\frac{\sqrt{\pi}}{2} H \left[ e^{-x^2} \right],$$
where $H[\cdot]$ denotes the Hilbert transform
$$H[f]=\frac{1}{\pi}{\rm p.v.}\int_{-\infty}^{\infty} \frac{f(y)}{x-y}dy.$$
As $\exp\left(-\frac{a}{2\nu}\xi^2\right)\int_0^\xi \exp\left(\frac{a}{2\nu}\eta^2\right) d\eta
=\sqrt{\frac{\pi\nu}{2a}} H\left[ \exp\left(-\frac{a}{2\nu}\xi^2\right)\right],$
we can write
$$U(\xi)=C'\sqrt{\frac{a}{2\pi\nu}} \exp\left(-\frac{a}{2\nu}\xi^2\right)
  + C H\left[\sqrt{\frac{a}{2\pi\nu}} \exp\left(-\frac{a}{2\nu}\xi^2\right)\right],$$
  with suitably redefined constants $C, C'$. This shows the general solution consists of
  two kinds of solutions, which we call a source-type solution and a kink-type one.
  The former converges to the Dirac mass and the latter to the Cauchy kernel $1/x$
  in the limit of $a/\nu \to \infty.$

\begin{figure}[ht]
\begin{minipage}{0.5\linewidth}
\includegraphics[scale=0.3,angle=0]{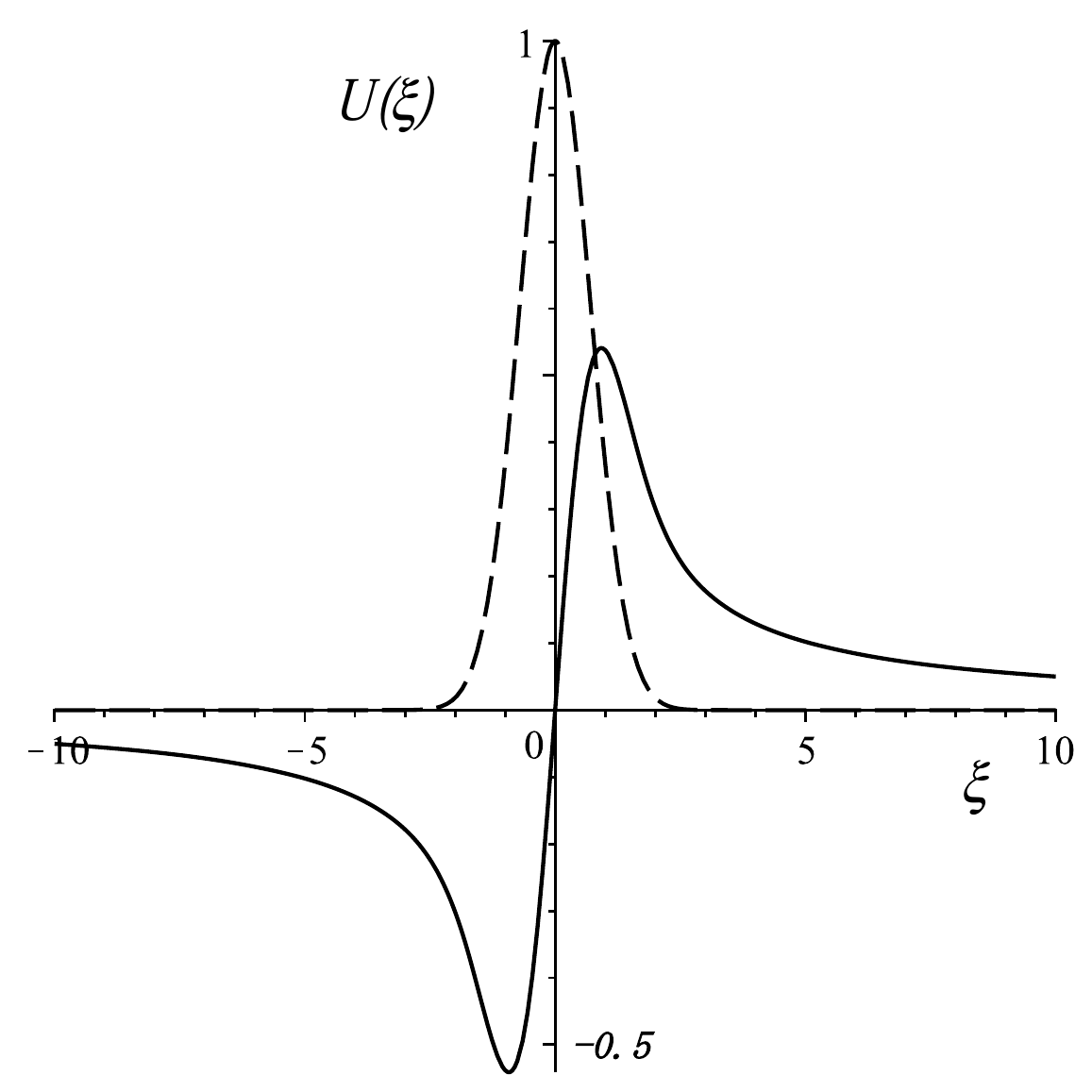}
\caption{Comparison of $\exp(-\xi^2)$ (dashed) with $D(\xi)=\frac{\sqrt{\pi}}{2}H[\exp(-\xi^2)]$ (solid). }
\label{Dawson1D}
\end{minipage}
\end{figure}

It is of interest to note that if a solution is found, then its Hilbert transform gives the other solution.
In fact applying $H[\cdot]$ to (\ref{1D_FP})
$$\partial_\xi^2 H[U]+\frac{a}{\nu}\partial_\xi (H[\xi U])=0.$$
By $H[\xi U]=\xi H[U]-\frac{1}{\pi}\int_{\mathbb{R}^1} U(\xi)d\xi,$ e.g. \cite{King2009},
it follows that
$$\partial_\xi^2 H[U]+\frac{a}{\nu}\partial_\xi (\xi H[U])=0,$$
which shows that $H[U]$ also a solves the same equation. See the Fig.\ref{Dawson1D}
for a comparison of those fundamental solutions.

\begin{table}
\begin{tabular}{|c|c|c|}
\hline
& initial data & non-Gaussian steady solutions \\\hline 
1D Burgers  & $u \propto \frac{1}{x} \in B^0_{1,\infty},\, \notin  L^1$ &  $u \propto x \in L^1_{\rm loc}$, continuous \\ \hline
2D Navier-Stokes  & ${\omega \propto \frac{1}{r^2} \in B^0_{1,\infty},\, \notin  L^1 \atop
\left( u \propto \frac{1}{r} \in B^0_{2,\infty},\,  \notin  L^2 \right)}$  & $\omega \propto \log r \in L^1_{\rm loc}$, discontinuous  \\\hline
3D Navier-Stokes  & ${\chi \propto \frac{1}{r^3} \in B^0_{1,\infty}, \notin  L^1 \atop
 \left( u \propto \frac{1}{r} \in B^0_{3,\infty},\,  \notin  L^3 \right)}$ &  $\chi \propto \frac{1}{r} \in L^1_{\rm loc}$, discontinuous\\
\hline
\end{tabular}
\caption{The behaviour of the initial data  and that of non-Gaussian steady solutions \textit{near the origin.}
  Note that the latter  behave just like Green's functions for the Poisson equations.}
\label{Near-origin}
\end{table}

\subsection{Two-dimensional case}
In the case of the 1D Burgers equation the singular self-similar initial condition becomes continuous
after infinitesimal time evolution. The second steady solution, which is not Gaussian, is also continuous
but not in $L^1(\mathbb{R}^1)$. We should take it into account when we discuss long-time evolution.

For the 2D (and 3D)  Navier-Stokes equations, the other steady solutions of non-Gaussian form, are in fact
neither continuous nor in $L^1$.
They are discontinuous at the origin.
Hence care should be taken in considering them when we discuss long-time evolution
in a larger function space such as $B^0_{1,\infty}$.

\begin{figure}[ht]
\begin{minipage}{0.5\linewidth}
  \includegraphics[scale=0.3,angle=0]{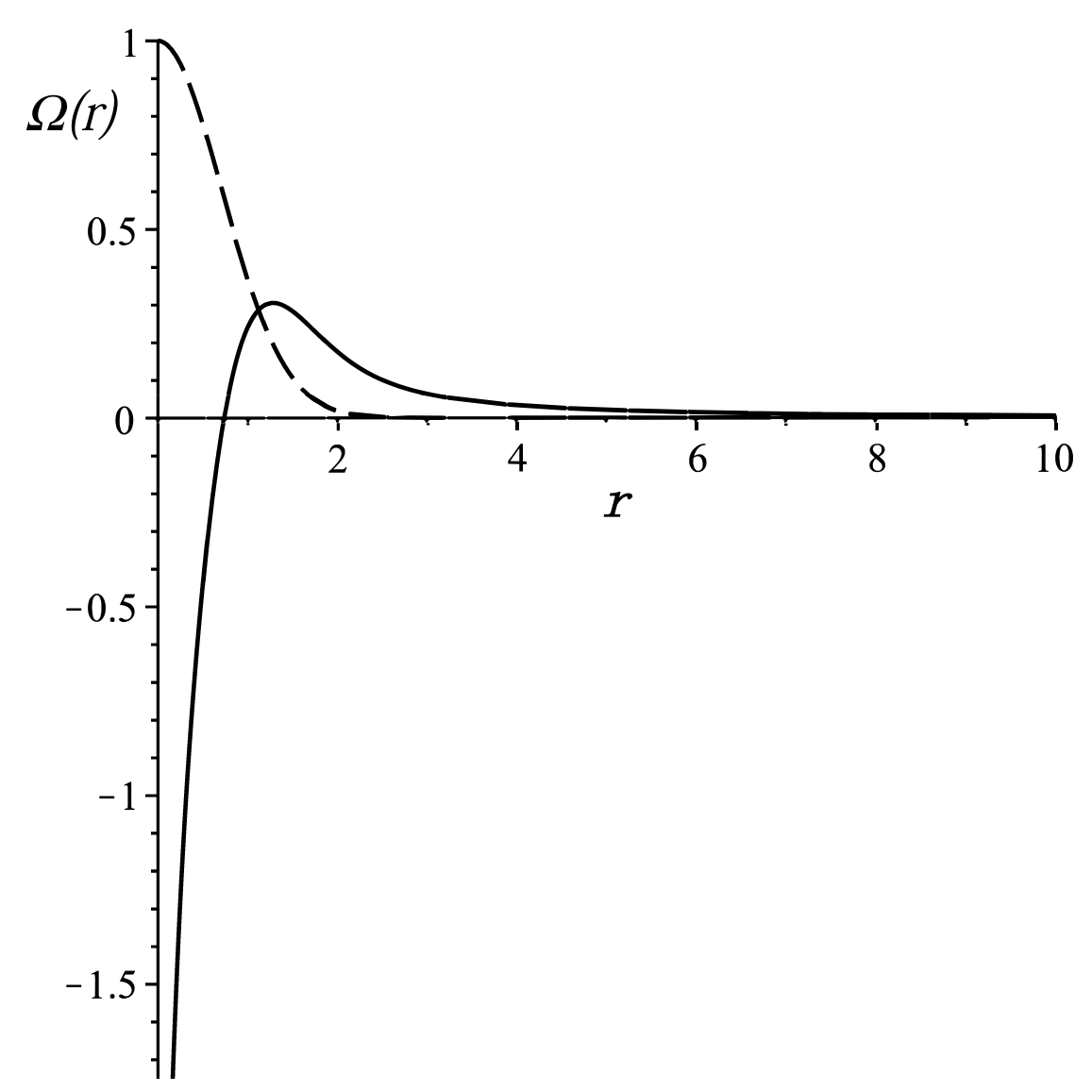}
\caption{Comparison of $\exp(-r^2)$ (dashed) with
  $\frac{1}{2}\exp(-r^2) ({\rm Ei}(r^2)-\gamma)$ (solid)  for $\frac{a}{2\nu}=1.$,
  where $\gamma=0.57721\ldots$ is the Euler's constant. The latter behaves as $1/r^2$
  as $r \to \infty$ and $\log r$   as $r \to 0+$.}
\label{Dawson2D}
\end{minipage}
\begin{minipage}{0.5\linewidth}
  \includegraphics[scale=0.3,angle=0]{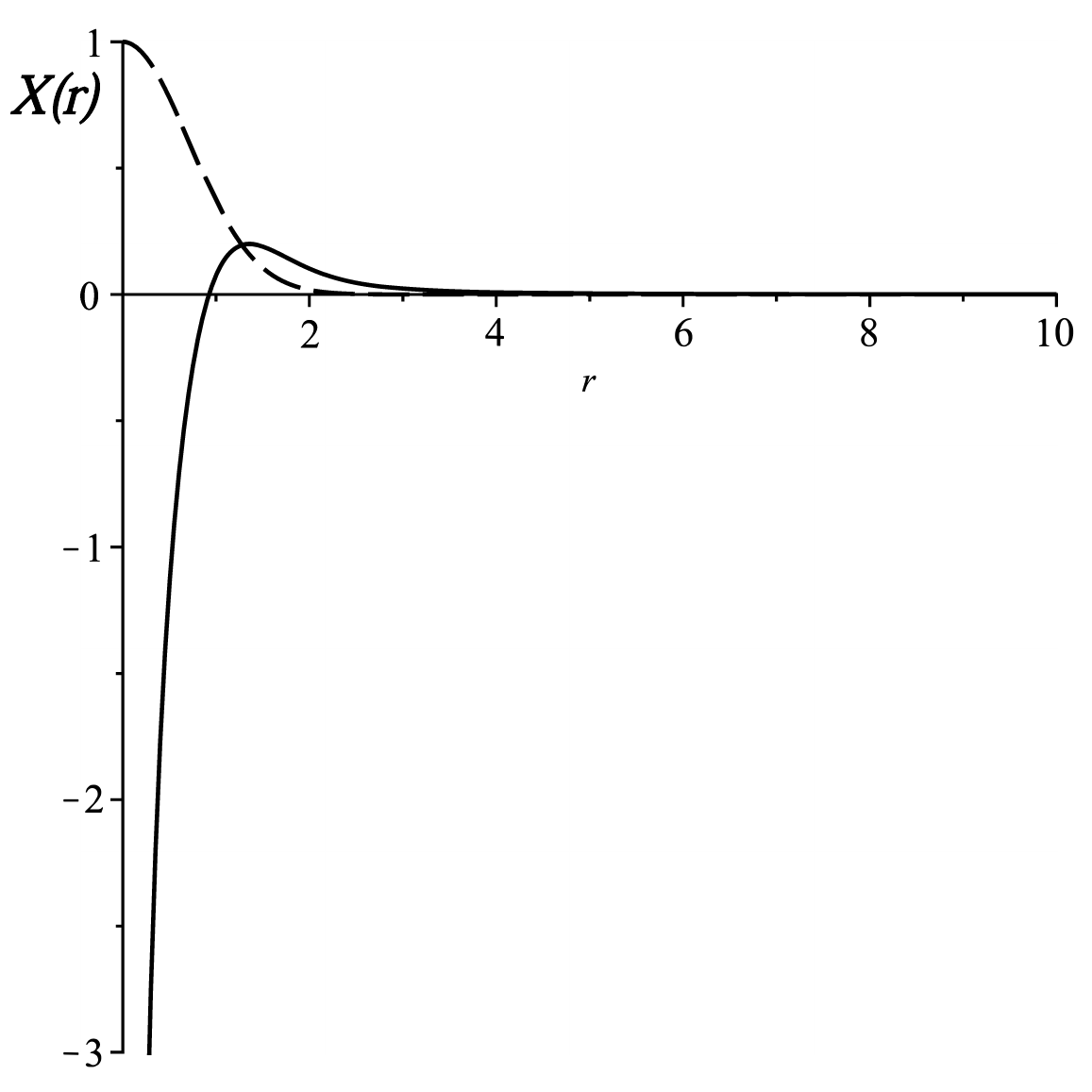}
  \caption{Comparison of $\exp(-r^2)$ (dashed) with $2D(r)-\frac{1}{r}$ (solid)
    for $\frac{a}{2\nu}=1.$
  The latter behaves $1/r^3$ as $r \to \infty$ and $-1/r$ as $r \to 0+$.}
  \label{Dawson3D}
\end{minipage}
\end{figure}

Under the assumption of radial symmetry
the Fokker-Planck equation in two dimensions takes the following form
$$\frac{1}{r}
\frac{d}{d r}\left(r \frac{d f}{d r}\right)
+\frac{a}{\nu r} \frac{d}{d r}
\left(r^2 \textcolor{black}{f}\right)=0,$$
which is equivalent to
$$\left(\frac{d}{d r}+\frac{1}{r}\right)\left(\frac{d f}{d r}
+\frac{a}{\nu} r f\right)=0.$$
Indeed,
$$f''+\frac{1}{r}f'+\frac{a}{\nu}\left(2f+r f'\right)=0,$$
$$ f''+\frac{a}{\nu}\left(f+r f'\right)
+\frac{1}{r}\left(f' +\frac{a}{\nu}rf \right)=0,$$
$$ \left(f'+\frac{a}{\nu}r f\right)'
+\frac{1}{r} ( f' +\frac{a}{\nu}rf ) =0.\;\;\square$$
Setting $h(r)=f' +\frac{a}{\nu}rf,$ we have $h(r)=C/r,$ that is,
$$\exp\left(-\frac{a}{2\nu}r^2 \right)
\left\{ f\exp\left(\frac{a}{2\nu}r^2\right)\right\}'=\frac{C}{r}.$$
Hence
\begin{eqnarray}
f&=& c_1 \exp\left(-\frac{a r^2}{2\nu} \right)
+c_2 \exp\left(-\frac{a r^2}{2\nu} \right)
{\rm f.p.} \int_0^r \exp\left(\frac{a s^2}{2\nu} \right)\frac{ds}{s}\nonumber\\
&=& c_1 \exp\left(-\frac{a r^2}{2\nu} \right)
+ \frac{c_2}{2} \exp\left(-\frac{a r^2}{2\nu} \right)
\left({\rm Ei}\left(\frac{ar^2}{2\nu}\right) -\log \left(\frac{a}{2\nu} \right) -\gamma  \right),\nonumber
\end{eqnarray}
where ${\rm Ei}(x)=-{\rm p.v.}\int_{-x}^{\infty} \frac{e^{-t}}{t} dt$ denotes the exponential integral and
$\gamma \approx 0.577$ the Euler's constant.

\color{black}
{\bf Derivation}\\
Define
$$I \equiv\lambda e^{-\lambda r^2}\int_\epsilon^r \frac{e^{-\lambda s^2}}{s} ds.$$
We have
$$\int_\epsilon^r \frac{e^{-\lambda s^2}}{s} ds
=\left[ \log s e^{-\lambda s^2}\right]_\epsilon^r -\int_\epsilon^r \log s\,2\lambda s e^{-\lambda s^2}ds$$
$$=e^{-\lambda r^2}\log r -\log \epsilon - \underbrace{2\lambda \int_\epsilon^r s \log s e^{-\lambda s^2}}_{=(A)}ds.$$
Now, with $u=e^{-\lambda s^2}, \epsilon'=e^{\lambda \epsilon^2}\approx 1$
$$(A)=\int_{e^{\lambda\epsilon^2}}^{e^{-\lambda r^2}}\log\left\{\frac{1}{\sqrt{\lambda}}(\log u)^{1/2} \right\}du$$
$$=\int_{e^{\lambda\epsilon^2}}^{e^{-\lambda r^2}} \log \frac{1}{\sqrt{\lambda}} du
+\frac{1}{2}\int_{e^{\lambda\epsilon^2}}^{e^{-\lambda r^2}}1 \log\log u du.$$
As
$$\int_{e^{\lambda\epsilon^2}}^{e^{-\lambda r^2}}1 \log\log u du
=\left[u\log\log u \right]_{e^{\lambda\epsilon^2}}^{e^{-\lambda r^2}}
-\int_{e^{\lambda\epsilon^2}}^{e^{-\lambda r^2}}\frac{du}{\log u}$$ 
$$\approx e^{\lambda r^2} \log(\lambda r^2)-\log \lambda -2 \log\epsilon -{\rm li}(e^{\lambda r^2})
+{\rm li}(e^{\lambda \epsilon^2}),$$
we find
$$(A)\approx\frac{1}{2}\log \lambda +e^{\lambda r^2}\log r -\frac{1}{2}{\rm li}(e^{\lambda r^2})+\frac{\gamma}{2}.$$
Hence
$$\int_\epsilon^r \frac{e^{-\lambda s^2}}{s} ds=
-\frac{1}{2}\log \lambda  \textcolor{black}{+}\frac{1}{2}{\rm li}(e^{\lambda r^2})-\frac{\gamma}{2}-\log \epsilon,$$
that is,
$$I=\lambda e^{-\lambda r^2}\int_\epsilon^r \frac{e^{-\lambda s^2}}{s} ds
=\frac{\lambda}{2} e^{-\lambda r^2}\left\{{\rm li}(e^{\lambda r^2})- \log \lambda -\gamma -2 \log \epsilon\right\}.$$
Taking its f.p.,
$$I=\frac{\lambda}{2} e^{-\lambda r^2}\left\{{\rm li}(e^{\lambda r^2})- \log \lambda -\gamma \right\}.$$
Here  $${\rm li}(x) \equiv {\rm p.v.} \int_0^x\frac{dt}{\log t},\;( x >1)$$ denotes the logarithmic integral
and
$${\rm Ei}(x) \approx \gamma +\log x +x \;\;\,\mbox{as}\;\; x \to 0+$$
$${\rm Ei}(x) \left(\equiv {\rm li}(e^x)\right)=O\left(\dfrac{e^x}{x}\right)\:\:\,\mbox{as}\;\; x \to \infty.$$
\color{black}
\subsection{Three-dimensional case}
Under the assumption of radial symmetry, the Fokker-Planck equation
$\triangle f+\frac{a}{\nu}\nabla \cdot(\bm{\xi}f)=0$
in three dimensions takes the following form 
$$\frac{1}{r^2}
\frac{d}{d r}\left(r^2 \frac{d f}{d r}\right)
+\frac{a}{\nu}\frac{1}{r^2} \frac{d}{d r}
\left(r^3 \textcolor{black}{f}\right)=0,$$
where $r=|\bm{\xi}|.$ It is equivalent to
$$\left(\frac{d}{d r}+\frac{2}{r}\right)\left(\frac{d f}{d r}
+\frac{a}{\nu} r f\right)=0.$$
Indeed,
$$f''+\frac{2}{r}f' +\frac{a}{\nu}(3f + r f')=0,$$
$$\left(f''+\frac{a}{\nu}(f+rf') \right)
+\left(\frac{2}{r}f'+\frac{2a}{\nu}f\right)=0,$$
$$\left(f'+\frac{a}{\nu}rf \right)'
+\frac{2}{r} ( f'+\frac{a}{\nu}rf )=0.\;\;\square$$
\noindent Setting $h(r)=\frac{d f}{d r}+\frac{a}{\nu} r f,$
we have $h(r)=C/r^2.$
Thus we can write
$$\exp\left(-\frac{a}{2\nu}r^2 \right)
\left\{ f\exp\left(\frac{a}{2\nu}r^2\right)\right\}'=\frac{C}{r^2}.$$
A further integration gives
$$f= c_1 \exp\left(-\frac{a r^2}{2\nu} \right)
+c_2 \exp\left(-\frac{a r^2}{2\nu} \right)
{\rm f.p.} \int_0^r \exp\left(\frac{a s^2}{2\nu} \right)\frac{ds}{s^2}$$
$$= c_1 \exp\left(-\frac{a r^2}{2\nu} \right)
+c_2 \left( \sqrt{\dfrac{2a}{\nu}} D\left(\sqrt{\frac{a}{2\nu}}r \right) -\frac{1}{r}\right),
$$
where $D(\cdot)$ denotes the Dawson's integral.

\color{black}
{\bf Derivation}\\
As
$$\int_\epsilon^r e^{\lambda s^2} \frac{ds}{s^2}
=\left[-\frac{e^{\lambda s^2}}{s} \right]_\epsilon^r
+2\lambda\int_\epsilon^r e^{\lambda s^2} ds,$$
we have
  $$ e^{-\lambda r^2}
  \int_\epsilon^r e^{\lambda s^2} \frac{ds}{s^2}
  =-\frac{1}{r}+\frac{e^{-\lambda r^2}}{\epsilon}
  +2\lambda e^{-\lambda r^2}\int_\epsilon^r e^{\lambda s^2} ds.$$
  Noting
  $$e^{-\lambda r^2}\int_0^r e^{\lambda s^2} ds=\frac{1}{\sqrt{\lambda}}D\left(\sqrt{\lambda}r\right)$$
  and dropping the  $O(\epsilon^{-1})$ term, we obtain the desired form.
\color{black}


\enlargethispage{20pt}





{\bf Acknowledgement}\\
    This work has been supported by EPSRC grant   EP/N022548/1.
Its major part  was carried out  when one of the authors (K.O.)
  was affiliated with School of Mathematics and Statistics, the University of Sheffield.



\vskip2pc


\end{document}